\documentclass[%
superscriptaddress,
twocolumn,
preprintnumbers,
nofootinbib,
 amsmath,amssymb,
 aps,
prd,
]{revtex4-2}

\usepackage{graphicx}
\usepackage{hyperref}
\usepackage{bm}

\usepackage{color}

\begin{document}

\preprint{UT-Komaba/22-4}

\title{Schwinger model on an interval: 
\\
analytic results and DMRG}

\author{Takuya Okuda}

\affiliation{Graduate School of Arts and Sciences, University of Tokyo\\
Komaba, Meguro-ku, Tokyo 153-8902, Japan} 

\email{takuya@hep1.c.u-tokyo.ac.jp}

\begin{abstract}
Quantum electrodynamics in $1+1$ dimensions (Schwinger model) on an interval admits lattice discretization with a finite-dimensional Hilbert space,  and is often used as a testbed for quantum and tensor network simulations.  In this work we clarify the precise mapping between the boundary conditions in the continuum and lattice theories. 
 In particular we show that the conventional Gauss law constraint commonly used in simulations induces a strong boundary effect on the charge density, reflecting the appearance of fractionalized charges.
Further, we obtain by bosonization a number of exact analytic results for local observables in the massless Schwinger model.  We compare these analytic results with the simulation results obtained by the density matrix renormalization group (DMRG) method and find excellent agreements.
\end{abstract}

\maketitle


\section{Introduction}

Quantum electrodynamics in $1+1$ dimensions, also known as the Schwinger model~\cite{Schwinger:1962tp}, is one of the simplest non-trivial gauge theories.
Since its introduction in the 60's it has been widely studied.
These days it is often used as a toy model to benchmark numerical techniques for quantum gauge theories, such as tensor network and quantum simulations.
See, for example,~\cite{Byrnes:2002nv,Banuls:2013jaa,Banuls:2013zva,Rico:2013qya,Buyens:2013yza,Banuls:2015sta,Banuls:2016lkq,Funcke:2019zna,Zache:2021ggw,Halimeh:2021ufh,Martinez:2016yna,Muschik:2016tws,Bernien_2017,Surace:2019dtp,Klco:2018kyo,Kokail:2018eiw,Magnifico:2019kyj,Yang:2020yer,Chakraborty:2020uhf,2020PhRvR...2b3342K,Yamamoto:2021vxp,deJong:2021wsd,2022Sci...377..311Z,Mildenberger:2022jqr}.

With the recent rapid development of quantum devices, quantum simulation of gauge theory is becoming more feasible.
For this purpose, as in classical simulation, we need to discretize the gauge theory and put it on a finite lattice.
In the noisy intermediate-scale quantum (NISQ) era~\cite{preskill2018quantum}, the number fo available qubits and the physical volume of the space on which the gauge theory is simulated will be limited.
For this reason, simple $(1+1)$-dimensional gauge theories such as the Schwinger model are natural targets of quantum simulation.
Putting these theories on a spatial interval rather than a circle has an advantage, because the Gauss law constraint allows us to remove gauge fields completely on an interval, while on a circle there remains an infinite-dimensional Hilbert space.
The spatial interval for the continuum model corresponds to the open boundary condition of the lattice model.
It is thus desirable to know the precise correspondence between the theories in the continuum and on the lattice.
To compare the continuum and lattice formulations, it also helps to have analytic results that take into account the strong effects of the boundaries and the finite volume.
Rather surprisingly, the study of such effects in the literature is limited.%
\footnote{%
See~\cite{Iso:1988zi,Sachs:1991en} for the study of the model on a circle with finite radius, and~\cite{Durr:1998ea,Kao:2001qn} for an earlier study of boundary effects.
}

With these motivations, in this paper we study the Schwinger model on a finite interval and clarify the precise mapping between the continuum (original and bosonized) and lattice models.
In particular, we show that the commonly used Gauss law constraint~\cite{Hamer:1997dx} in the lattice formulation induces fractionalized charges on the boundaries, and demonstrate that for an alternative constraint~\cite{Berruto:1997jv} the boundary charges are also modified.%
\footnote{%
If the periodic boundary condition is chosen, the modification of the Gauss law is equivalent to the mass shift studied in~\cite{Dempsey:2022nys} via a field redefinition.
}
Along the way we establish the precise correspondence between the boundary conditions in different formulations.
We also derive a number of analytic expressions for physical observables in the ground state in the massless case.
This is possible because bosonization maps the massless Schwinger model to a free scalar theory~\cite{Coleman:1975pw,Coleman:1976uz}.
Some of these analytic results were used in~\cite{q-sim-confinement} to compare with the results of digital quantum simulation of the lattice Schwinger model on a classical simulator.

The paper is organized as follows.
In Section~\ref{sec:continuous} we review the continuum Schwinger model in the original formulation.
In Section~\ref{sec:bosonization} we study the Schwinger model on an interval using bosonization and derive some analytic results.
Section~\ref{sec:lattice} contains our study of the Kogut-Susskind lattice formulation of the Schwinger model on a finite lattice with the open boundary condition.
We review two equivalent formulations, one based on the staggered fermion and another based on spin variables.
We compute by the density matrix renormalization group (DMRG)~\cite{White:1992zz,PhysRevB.48.10345} some physical observables in the ground state and find agreement with the analytic results from Section~\ref{sec:bosonization}, using the original and modified Gauss law constraints.
We conclude the paper with discussion in Section~\ref{sec:conclusion}.
In Appendix~\ref{app:images} we calculate the energy in the presence of probe charges using the method of images.
In Appendix~\ref{sec:one-form}  we show that the general lattice QCD in the Kogut-Susskind formulation~\cite{Kogut:1974ag} enjoys an exact one-form symmetry for the part of the center of the gauge group under which the matter fermions are neutral.

\section{Continuum Schwinger model on an interval}\label{sec:continuous}

In this section we study the continuum Schwinger model on an interval.
We first review the original fermionic formulation of the model.
Then we review the bosonized version and derive a number of new analytic results for local observables.

\subsection{Review of the fermionic formulation}
\label{sec:fermionic-formulation}

We use notations $x^0=t$, $x^1=x$ for spacetime coordinates and use the Minkowski metric $\eta_{\mu\nu}=\text{diag}(1,-1)$ to raise and lower indices.
The dynamical fields in the Schwinger model are the gauge field $A_\mu$ ($\mu=0,1$) and the Dirac fermion $\psi=(\psi_u,\psi_d)^T$ which is a two-component spinor.
Let $g$ be the gauge coupling and $m$ the fermion mass.
The model is defined by the action
\begin{equation}\label{eq:Schwinger-action}
\begin{aligned}
S
&=\int d^2x \Big[ -\frac{1}{4} F_{\mu\nu} F^{\mu\nu} +\frac{g\Theta(x)}{4\pi} \epsilon_{\mu\nu} F^{\mu\nu}
\\
&
 +{\rm i}\bar{\psi}\gamma^\mu (\partial_\mu +{\rm i} g A_\mu ) \psi 
-m\bar{\psi}\psi  \Big] +\text{bdry terms}
 \,.
\end{aligned}
\end{equation}
We use the notations
\begin{equation}
\epsilon_{01}=-\epsilon^{01}=1\,,\
\gamma^0 = \sigma^3\,,\
\gamma^1 = {\rm i}\sigma^2 \,, \
\gamma^5 = \gamma^0\gamma^1 \,,
\end{equation}
and 
$\bar\psi=\psi^\dagger\gamma^0$.
We allow 
the theta angle
to be position-dependent and denote it by $\Theta(x)$.

\begin{figure}[t]
\begin{center}
\includegraphics[scale=.28]{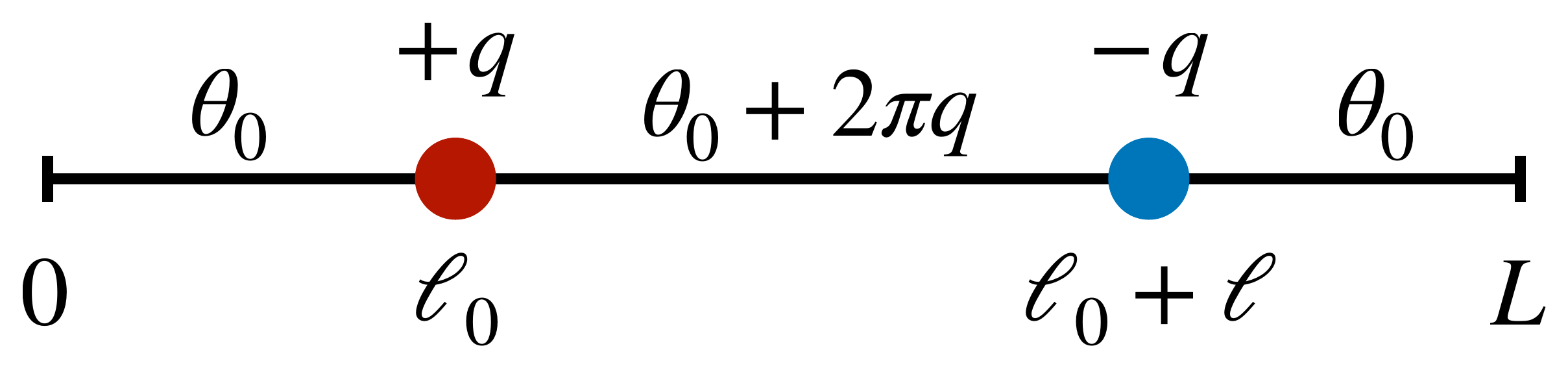}
\end{center}
\caption{The setup for $\Theta(x)=\Theta_{(q,\theta_0)} $ in~(\ref{eq:Theta-x}), corresponding to probe charges $+q$ at $x=\ell_0 $ and $-q$ at $x=\ell_0+\ell $.
}\label{fig:probes-interval}
\end{figure}

Consider, for example,
\begin{equation}\label{eq:Theta-x}
\Theta_{(q,\theta_0)} (x) =\left \{
\begin{array}{ccc}
\displaystyle  \theta_0 + 2\pi q & \text{ for } & \displaystyle \ell_0  < x <  \ell_0+\ell \,.\\
\displaystyle \theta_0 & \text{ for }  & \text{otherwise} \,.
\end{array}
\right.
\end{equation}
See FIG.~\ref{fig:probes-interval}.
The discrete changes in the theta angle $\Theta(x)$ correspond to the presence of probe charges.
Indeed we can rewrite the relevant part of the action as
\begin{align}
&\quad
\int d^2x \frac{\Theta_{(q,\theta_0)} }{4\pi} \epsilon_{\mu\nu} F^{\mu\nu} 
\nonumber\\
&=
\int d^2x \left[
\frac{\theta_0}{4\pi} \epsilon_{\mu\nu} F^{\mu\nu} 
\hspace{-1mm}
- q \left[ \delta(x-\ell_0) -  \delta(x-\ell_0-\ell)\right] A_0
\right] \,, \nonumber
\end{align}
where we explicitly see the point-like sources for the gauge field.

Let us study the model on an interval~$0\leq x\leq L$.
For the fermion~$\psi$, the general boundary conditions (b.c.'s) at each boundary, compatible with the variational principle, are parametrized by a real parameter $\nu$ mod~$\mathbb{Z}$:%
\footnote{%
These b.c.'s  preserve (explicitly break) the vector (axial) $U(1)$ symmetry generated by $\bar\psi\gamma^\mu\psi$
($\bar\psi\gamma^\mu\gamma^5\psi$).
}
\begin{equation}\label{eq:nu-def}
\psi_{\rm L}  + e^{2 \pi {\rm i}\nu}\psi_{\rm R} =0\,,
\end{equation}
where we defined $\psi_{\rm L}:= (\psi_u+\psi_d)/2$, $\psi_{\rm R}:= (\psi_u-\psi_d)/2$.
We are particularly interested in $(\psi_u,\psi_d,\nu)=(0,{\rm arbitrary},0), ({\rm arbitrary},0,\pi)$.
%
Up to a field redefinition $\psi\rightarrow \gamma^5\psi$, there are two inequivalent choices~\cite{Polchinski:1998rr}:
the Ramond (R) b.c.
\begin{equation}
 \psi_L(0)= s\, \psi_R(0) \text{ and }  \psi_L(L)= s\, \psi_R(L) 
\end{equation}
and the Neveu-Schwarz (NS) b.c.
 \begin{equation}
 \psi_L(0)= s\, \psi_R(0) \text{ and }  \psi_L(L)=- s \,\psi_R(L) \,,
\end{equation}
with $s=\pm 1$.
\footnote{%
Let us extend the domain
 of $\psi_{\rm L}(x)$ to $[-L,L]$ by 
$\psi_{\rm L}(x):=
-e^{2\pi {\rm i}\nu_0}
 \psi_{\rm R}(-x)$ for $-L\leq x\leq 0$.
 Since $\psi_{\rm L}(L)=e^{2\pi {\rm i}(\nu_1-\nu_0)} \times \psi_{\rm L}(-L)$,
$\psi_L$ 
 is periodic (anti-periodic) for the R (NS) b.c.
}

We work in the temporal gauge $A_0=0$, where the Gauss law constraint $\delta S/\delta A_0=0$ should be imposed on physical states.
Varying $A_0$, we find
 the Gauss law 
\begin{equation}\label{eq:Gauss-continuum}
 \partial_1 F_{01} =  \frac{g}{2\pi} \partial_1\Theta + g  \bar\psi\gamma^0\psi  
\end{equation}
in the bulk.
Composite operators such as~$\bar\psi\gamma^0\psi$ should be defined by some normal ordering~\cite{Casher:1974vf}.
Throughout this paper we make this implicit and omit normal ordering symbols for the fermion.
We will specify the b.c.'s on~$F_{01}$ at $x=0$ and $x=L$ in Section~\ref{sec:bosonization} where we study the continuum model in the bosonized formulation.
The boundary terms in~(\ref{eq:Schwinger-action}), which we do not write explicitly, should be chosen so that they are compatible with the b.c.'s.

The canonical momentum conjugate to $A^1$($=-A_1$) is 
\begin{equation}
\Pi=\partial_0 A^1 +\frac{g}{2\pi}\Theta \,.
\end{equation}
The density $\mathcal{H}$ of the Hamiltonian
$H=\int_0^L dx \, \mathcal{H}(x)$
is 
\begin{equation}\label{eq:Hamiltonian-density}
\mathcal{H}(x)=\frac{1}{2} \left( \Pi -\frac{g\Theta(x)}{2\pi} \right)^2
-{\rm i} \bar{\psi} \gamma^1 (\partial_1 + {\rm i} g A_1) \psi +m\bar{\psi} \psi  \,.
\nonumber
\end{equation}

Let us denote the expectation value of the operator $\mathcal{O}$ in the ground state by $\langle\mathcal{O}\rangle$. 
Local observables of the continuum Schwinger model on an interval include the energy density~$\langle \mathcal{H} \rangle$, the charge density~$\langle \bar\psi\gamma^0\psi\rangle$, the chiral condensate~$\langle \bar\psi\psi\rangle$, and the electric field $F_{01}$.

\subsection{Bosonized Schwinger model}
\label{sec:bosonization}

In this subsection we study the Schwinger model in the bosonized formulation.
There is some overlap with the appendix of~\cite{q-sim-confinement} that uses the same convention, and we refer the reader to that paper for details omitted here.

The bosonized Lagrangian density is~(cf. \cite{Gross:1995bp})
\begin{align}
\mathcal{L}&=
-\frac14 F_{\mu\nu}F^{\mu\nu} 
+\frac{g}{4\pi}\Theta(x^1) \epsilon^{\mu\nu}F_{\mu\nu} 
+ \frac{g}{\sqrt\pi} \epsilon^{\mu\nu} A_\mu \partial_\nu \phi 
\nonumber\\
&\qquad + \frac12 \partial_\mu\phi\partial^\mu\phi 
+ m g \frac{{\rm e}^\gamma}{2\pi^{3/2}}  \cos(2\sqrt\pi\phi) \,.
\end{align}
We choose an appropriate boundary condition on the gauge field so that the solution to the Gauss law constraint is
\begin{equation}\label{eq:Gauss-solution-bosonized}
F_{01} -\frac{g}{2\pi}\Theta = \frac{g}{\sqrt\pi}\phi \,.
\end{equation}
The Hamiltonian density  is given as 
\begin{align} 
\mathcal{H} & :=
\frac12(\Pi_\phi)^2+\frac12(\partial_x\phi)^2 
+
\frac{\mu^2}{2}
\left(\phi + \frac{\Theta(x)}{2\sqrt\pi}\right)^2
\nonumber\\
&\qquad
- mg \frac{{\rm e}^\gamma}{2\pi^{3/2}} :\hspace{-2pt} \cos(2\sqrt\pi \phi)\hspace{-2pt}:_\infty \,,
\label{eq:Hamiltonian-density-bosonized}
\end{align}
where $\Pi_\phi$ is the canonical momentum conjugate to $\phi$
and $\mu\equiv g/\sqrt\pi$.
We write $ :\hspace{-2pt} \bullet \hspace{-2pt}:_\infty$ for the normal ordering (see below) with respect to the creation-annihilation operators defined
in the infinite volume and used the
 relation
\begin{equation}
\bar\psi\psi =  - \displaystyle\frac{{\rm e}^\gamma}{2\pi^{3/2}} g   :\hspace{-2pt} \cos[2\sqrt\pi \phi(x)] \hspace{-2pt}:_\infty  \,,
\end{equation}
where $\gamma\simeq 0.58$ is the Euler constant.
The particular numerical coefficient ${\rm e}^\gamma/(2\pi^{3/2})$ is correct for this choice of normal ordering.%
\footnote{See~\cite{Coleman:1974bu} for a general discussion of normal ordering.}

We study the bosonized model with $m=0$ and the Dirichlet boundary conditions
\begin{equation} \label{def:w0-w1}
\phi =\sqrt\pi w_0 
 \text{ at } x=0,
\quad
\phi =\sqrt\pi w_1
 \text{ at } x=L .
\end{equation}
We set $k_n:= \pi n/L$.
Let us define 
\begin{equation}
\phi_0(x):= \sqrt\pi w_0+\sqrt\pi (w_1-w_0)\frac{x}{L}\,,
\end{equation}
\begin{equation}\label{eq:Theta-hat-def}
  \hat  \phi(x):=\phi(x)-\phi_0(x)\,,
  \quad
      \hat\Theta(x) := \Theta(x)+2\sqrt\pi \phi_0(x).
\end{equation}
Let us consider the Fourier expansions
\begin{equation}
\begin{aligned}
&\Pi_\phi (x)  = \sum_{n=1}^\infty \Pi_n \sin \left(k_n x\right) 
\,,\
\hat\phi (x)  = \sum_{n=1}^\infty \phi_n \sin\left(k_n x\right)  \,,
\\
&
\qquad\qquad\qquad\quad
\hat \Theta(x)   = \sum_{n=1}^\infty \Theta_n \sin\left(k_n x\right) \,.
\end{aligned}
\nonumber
\end{equation}
The Hamiltonian becomes
\begin{align}
&H_{\rm boson}=\frac{\pi(w_1-w_0)^2}{2L}
\nonumber \\
&\qquad \qquad 
+\sum_{n=1}^\infty 
\Big[
\omega_n \left(a_n^\dagger a_n
+\frac12\right) 
+
\frac{L\mu^2}{16}
\frac{k_n^2}{\omega_n^2}\Theta_n^2
\Big]\,,
\label{eq:Hamiltonian-bosonized}
\end{align}
where $\omega_n =  \sqrt{\mu^2 + k_n^2}$ and
\begin{equation} \label{interval-an}
a_n = \frac{\sqrt{L\omega_n}}{2}\left(\phi_n + \frac{\mu^2 \Theta _n}{2 \sqrt \pi \omega_n^2}\right)+ \frac{\rm i}{2} \sqrt{\frac{L}{\omega_n}} \Pi_n   \,.
\end{equation}
We have $[a_n,a_{n'}^\dagger ] = \delta_{nn'}$.
The ground state $|0\rangle$ satisfies $a_n |0\rangle=0$ and has a divergent energy due to the terms proportional to $\omega_n$, which are independent of $\Theta$.

The energy density $\langle 0| \mathcal{H}(x) |0\rangle$ is also UV divergent.
Let $\lfloor x\rfloor$ denote the largest integer smaller than or equal to $x$.
With a cut-off $k_n\leq \Lambda$,%
\footnote{%
For plots throughout the paper, we use Mathematica to evaluate regularized sums numerically by setting $[L\Lambda/\pi]$ to $10^4$.
} 
the regularized energy density is
\begin{align}
\mathcal{E}_\Lambda(x)&=
\frac{1}{2L}
\hspace{-2mm}
\sum_{n=1}^{\lfloor L\Lambda/\pi\rfloor}  
\hspace{-1mm}
\left[
\hspace{-1mm}
 \left(\omega_n +\frac{\mu^2}{\omega_n} \right) \sin^2 (k_n x)
 \hspace{-0.5mm}
+
\hspace{-0.5mm}
 \frac{ k_n^2}{\omega_n}\cos^2 (k_n x)\right]
\nonumber \\
&\quad
+\frac1{8\pi}
\left[
\left(\sum_{n=1}^{\lfloor L\Lambda/\pi\rfloor} \frac{\mu^2 k_n}{ \omega_n^2}\Theta _n\cos (k_n x)
\right)^2
\right.
\nonumber \\
&\quad\quad
+
\left.
 \left(\sum_{n=1}^{\lfloor L\Lambda/\pi\rfloor}  \frac{\mu k_n^2}{\omega_n^2} \Theta_n
 \sin  (k_n x)\right)^2
 \right] \,,\label{eq:energy-density-interval}
\end{align}
which is quadratically divergent. 
On a full infinite line without probe charges, the corresponding regularized energy density is, with $\omega(k):=\sqrt{k^2+\mu^2}$,%
\footnote{%
Explicitly, $\mathcal{E}^\text{line}_\Lambda= \left[\left(\Lambda^2+ \mu^2\right)^{1/2} \Lambda+ \mu^2 \sinh ^{-1}(\Lambda/\mu)\right]/4 \pi$.
}
\begin{equation}
\mathcal{E}^\text{line}_\Lambda:=
 \lim_{L\rightarrow \infty} \mathcal{E}_\Lambda(L/2) = 
 \int_0^\Lambda\frac{dk}{2\pi}\omega(k) \,.
\end{equation}
We define the renormalized energy density as
\begin{equation}\label{eq:energy-density-renormalized}
\mathcal{E}(x):= \lim_{\Lambda\rightarrow\infty}\left( \mathcal{E}_\Lambda(x) - \mathcal{E}^\text{line}_\Lambda\right) \,.
\end{equation}
An expression for the chiral condensate was found in~\cite{q-sim-confinement}:
\begin{equation}\label{eq:chiral-condensate-Theta}
\begin{aligned}
\langle \bar\psi\psi (x)\rangle
&=
- \displaystyle\frac{{\rm e}^\gamma g}{2\pi^{3/2}} 
\lambda(x)
\\
&\
\times\cos\bigg[ 2 \sqrt \pi \phi_0(x)
\hspace{-0.5mm}
-
\hspace{-0.5mm}
\sum_{n=1}^\infty \frac{\mu^2}{\omega_n^2} \Theta_n  \sin(k_n x)
\bigg] \,, 
\end{aligned}
\end{equation}
where%
\footnote{%
By several manipulations, one may rewrite (\ref{def:lambda}) as 
\begin{equation}
\hspace{-1mm}
\log \lambda(x)  
= 
\hspace{-1mm}
  \int_1^\infty
  \hspace{-3mm}
   \frac{du}{\sqrt{u^2-1}} 
   \hspace{-1mm}
 \left (\frac{-2}{{\rm e}^{2 \mu L  u}-1}
 \hspace{-0.5mm}
 +
 \hspace{-0.5mm}
\frac{\cosh\left[(2 x/L - 1)\mu L  u\right]}{\sinh [ \mu L u]} \right) \,.
\end{equation}
}
\begin{equation}\label{def:lambda}
\hspace{-1mm}
\lambda(x) := 
\hspace{-1mm}
\lim_{\Lambda\rightarrow \infty} 
\hspace{-1mm}
\exp
\hspace{-1mm}
\Bigg[  \sinh^{-1}
\hspace{-1mm}
\left(
\frac{\Lambda}{\mu}\right) 
-
\hspace{-3mm}
\sum_{n=1}^{\lfloor L\Lambda/\pi\rfloor}
\hspace{-1mm}
 \frac{2\pi}{L} \frac{\sin^2(k_nx)}{\sqrt{\mu^2+ k_n^2}}
\Bigg]\,.
\hspace{-1mm}
\end{equation}
For the charge density $ \bar\psi\gamma^0\psi (x)=   \partial_x \phi/\sqrt\pi$, we obtain
\begin{align}
\langle \bar\psi\gamma^0\psi (x)\rangle
=\frac{w_1-w_0}{L}  -\frac{\mu^2}{2\pi} \sum_{n=1}^\infty \frac{k_n}{\omega_n^2} \Theta_n \cos(k_n x)  \,.
\label{eq:charge-density-interval}
\end{align}
For the electric field we have
\begin{equation}\label{eq:electric-field-interval}
 \left\langle F_{01}  \right\rangle
= \frac{g}{2\pi} \sum_{n=1}^\infty \frac{k_n^2}{\omega_n^2} \Theta_n \sin(k_n x)
 \,.
\end{equation}
Below, we consider special and limiting cases.

\begin{figure*}[t]
\begin{center}
\begin{tabular}{cc}
\includegraphics[scale=.45]{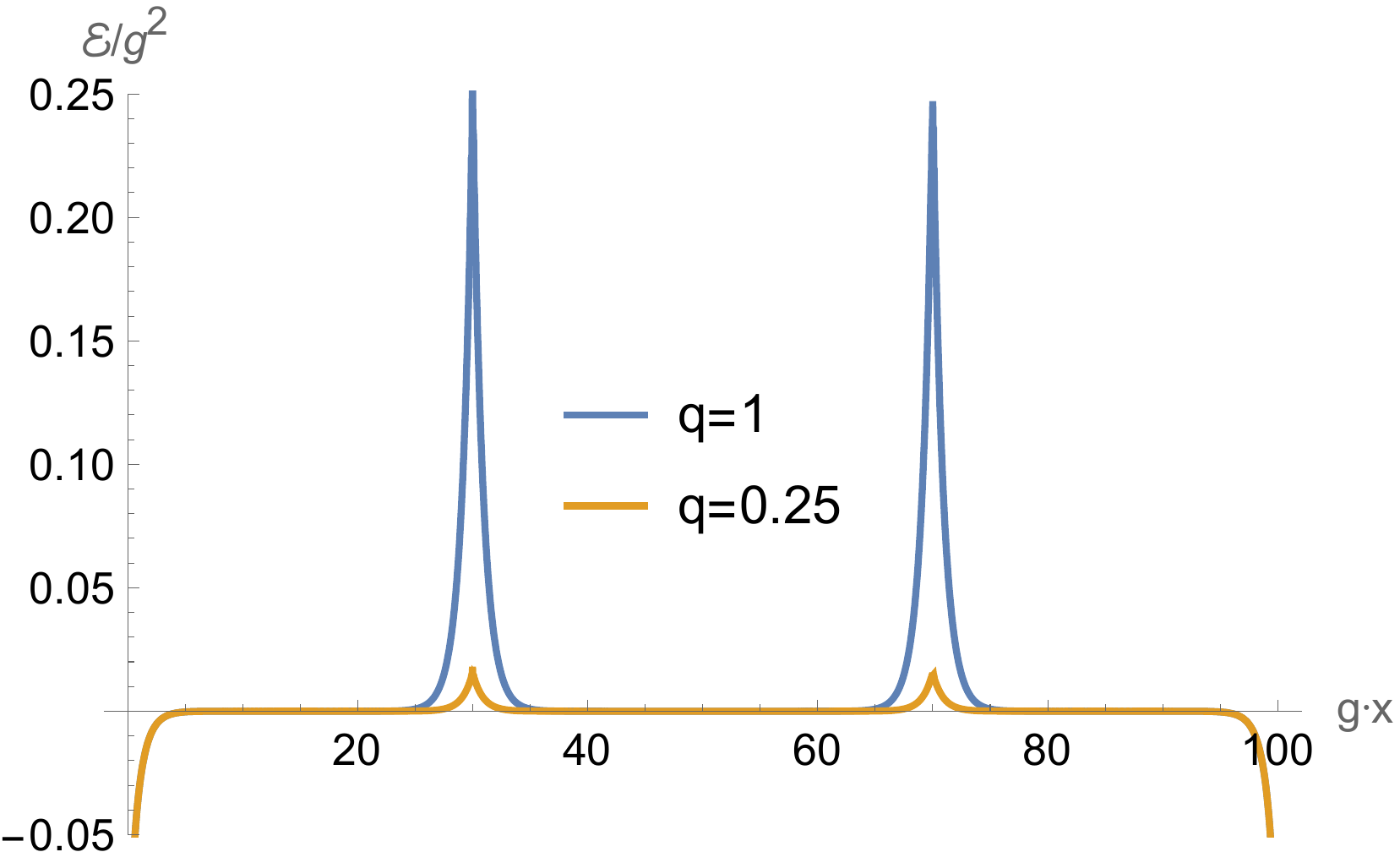}
&
\hspace{2mm}
\includegraphics[scale=.45]{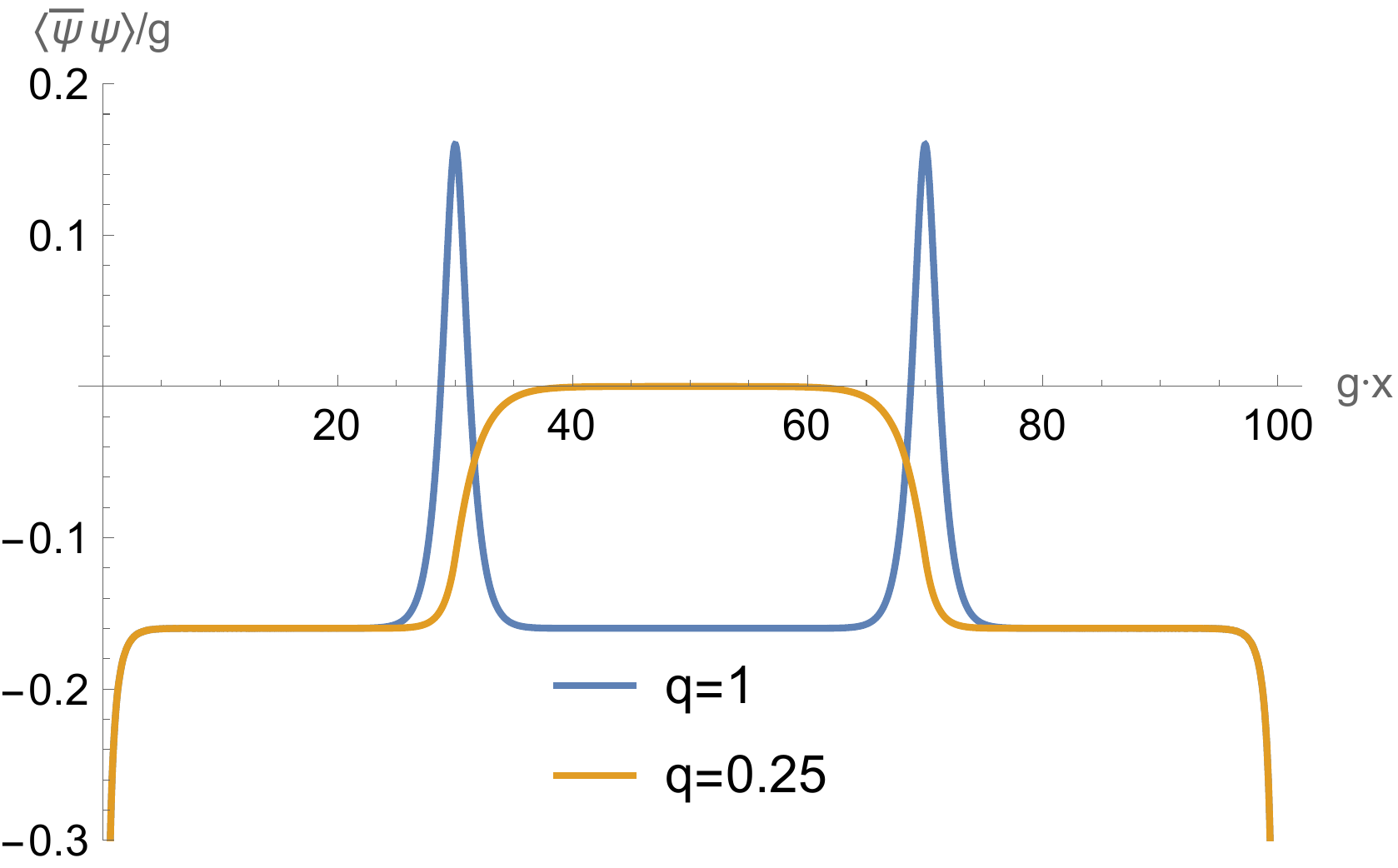}
\\
(a) & 
\hspace{2mm}(b)
\end{tabular}
\end{center}
\caption{%
(a) Renormalized energy density $\mathcal{E}$ given by~(\ref{eq:energy-density-renormalized}), for two probe charges $\pm q$ placed at $x= (L\mp \ell)/2$ represented by the position-dependent theta angle~(\ref{eq:Theta-x-2-probes}), for $L=100 g^{-1}$ and $\ell=40 g^{-1}$.
The local behaviors near each boundary and each pole are given by~(\ref{eq:energy-density-half-line}) and~(\ref{eq:energy-density-probe}), respectively.
(b) Chiral condensate $\langle \bar\psi\psi(x)\rangle$ given by~(\ref{eq:chiral-condensate-Theta}) for the same set-up.
The local behaviors near each boundary and each pole are given by~(\ref{eq:condensate-half-line}) and~(\ref{eq:condensate-probe}), respectively.
}\label{figure:energy-density-condensate-graph}
\end{figure*}

\paragraph{Two probe charges on an interval.}

\begin{figure*}[t]
\begin{center}
\begin{tabular}{cc}
\includegraphics[scale=.5]{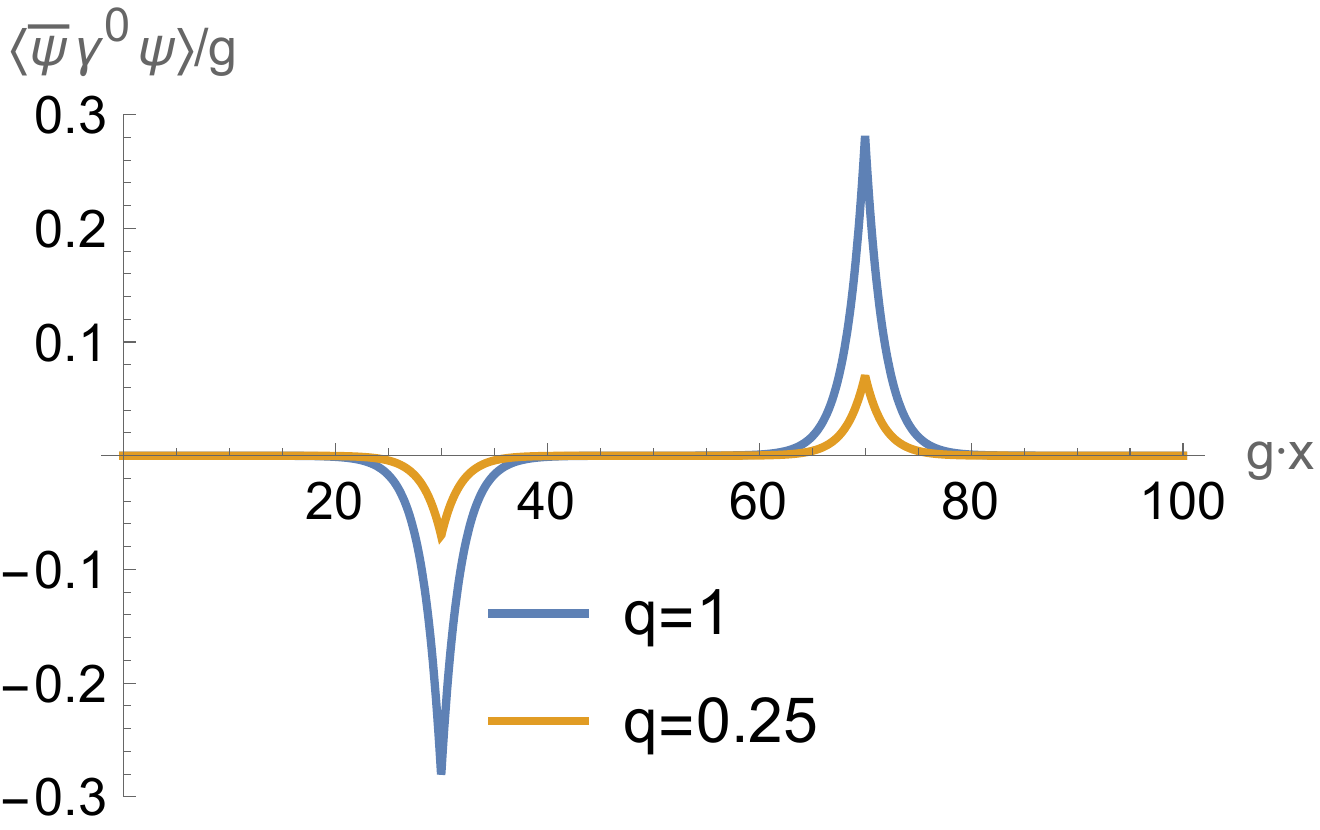} 
&
\hspace{2mm}
\includegraphics[scale=.71]{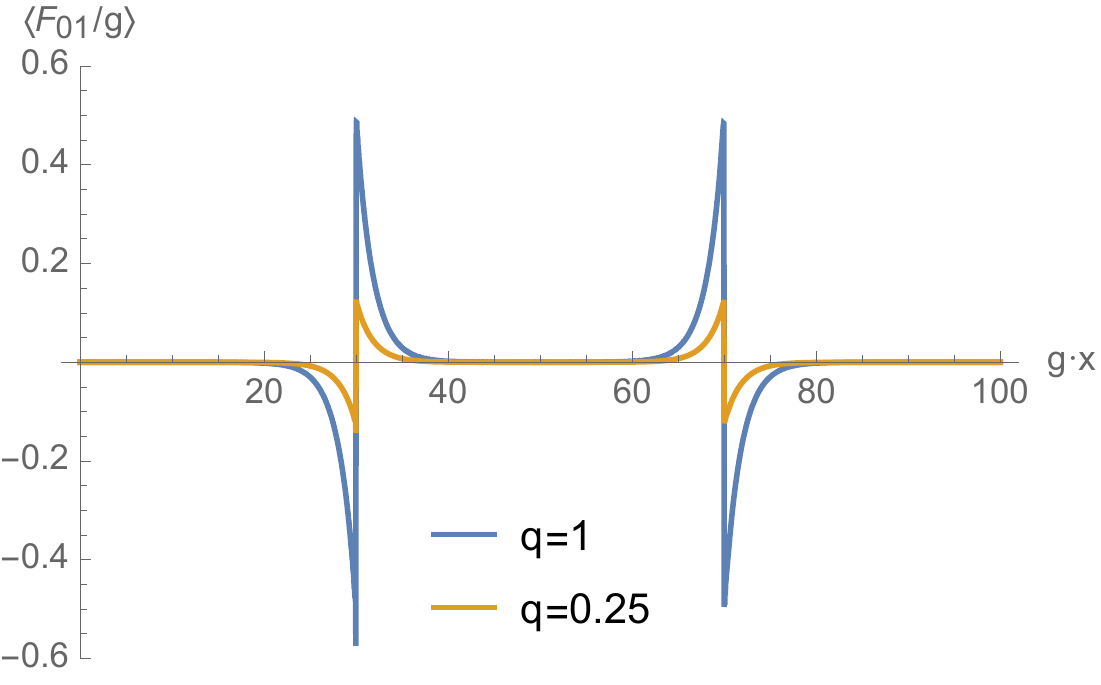}
\\
(a)  & 
\hspace{2mm}
(b)
\end{tabular}
\end{center}
\caption{%
(a) Charge density $\langle \bar\psi\gamma^0\psi(x)\rangle$ given by~(\ref{eq:charge-density-interval}) or equivalently by~(\ref{eq:eq:charge-density-pair})
for the same set-up as for FIG.~(\ref{figure:energy-density-condensate-graph}).
The local behaviors near the probes are given by~(\ref{eq:charge-density-probe}).
(b) Electric field $F_{01}$ given by~(\ref{eq:electric-field-interval}) or equivalently by~(\ref{eq:electric-field-pair}) for the same set-up.
}\label{figure:charge-density-electric-field-graph}
\end{figure*}

For probe charges on an interval, 
one can evaluate the sums above.
As an example, let us consider 
\begin{equation}\label{eq:Theta-x-2-probes}
\Theta_{\rm pair}(x) =\left \{
\begin{array}{cc}
\displaystyle 2\pi q & \displaystyle \text{for }   \frac{L-\ell}{2}\leq x\leq  \frac{L+\ell}{2} ,\\
\displaystyle 0 & \text{otherwise,}
\end{array}
\right.
\end{equation}
which represent a pair of charges $\pm q$ placed at $x= (L\mp \ell)/2$, {\it i.e.}, 
$\Theta_{\rm pair} = \Theta_{q,\theta_0=0}|_{\ell_0=(L-\ell)/2}$.
We impose the boundary conditions $\phi=0$ at $x=0,L$ corresponding to $w_0=w_1=0$.
The non-zero Fourier coefficients are
\begin{equation}
(\Theta_{\rm pair})_{2j+1} = 
 \frac{8q}{2j+1}  (-1)^{j} \sin \left[ k_{2j+1} (\ell/2)\right] 
\end{equation}
for $j \in \mathbb{Z}_{>0}$.
The total energy $E_{\rm pair}$, defined as the 
energy computed from~(\ref{eq:Hamiltonian-bosonized}) by removing terms proportional to $\omega_n$,
was obtained in~\cite{q-sim-confinement}:%
\footnote{%
In Appendix~\ref{app:images}, we give an alternative derivation of~(\ref{eq:E-pair}) by the method of images.
}
\begin{equation} \label{eq:E-pair}
E_{\rm pair} = 
\frac{\sqrt\pi}{2}
q^2 g \frac{(1-{\rm e}^{-\mu \ell})(1+ {\rm e}^{-\mu (L-\ell)})}{1+{\rm e}^{-\mu L}} \,.
\end{equation}
The energy density~$\mathcal{E}(x)$ in~(\ref{eq:energy-density-renormalized}) computed for (\ref{eq:Theta-x-2-probes}) is plotted in FIG.~\ref{figure:energy-density-condensate-graph}(a).%
\footnote{%
It is possible to perform the summations in~(\ref{eq:energy-density-interval}) to obtain an alternative expression.
We found the result to be not illuminating.
}
The chiral condensate~$\langle \bar\psi\psi (x)\rangle$ in~(\ref{eq:chiral-condensate-Theta}) corresponding to~(\ref{eq:Theta-x-2-probes}) is plotted in FIG.~\ref{figure:energy-density-condensate-graph}(b).
For the cosine in%
~(\ref{eq:chiral-condensate-Theta}),
the residue method gives its argument explicitly as
\begin{equation}
-\pi q  \sum_{j=0,1}
 (-1)^{\lfloor \eta _j(x) \rfloor}
\left(
\frac{ \cosh[ (  \{\eta_+(x)\}  -1/2)  \mu L]}{\cosh (\mu L/2)}  -1\right) \,,
\nonumber
 \end{equation} 
 where $\eta_0(x): = (x-  \frac{L-\ell}{2})/L$, $\eta_1(x): = (x-  \frac{L+\ell}{2})/L$, and $\{\eta\} := \eta - \lfloor \eta \rfloor$.
The charge density~$\langle \bar\psi\gamma^0\psi (x)\rangle$
 that corresponds to (\ref{eq:Theta-x-2-probes}) is plotted in FIG.~\ref{figure:charge-density-electric-field-graph}(a).
The summation in~(\ref{eq:charge-density-interval}) can be performed explicitly to give
\begin{align}\label{eq:eq:charge-density-pair}
\langle \bar\psi\gamma^0\psi (x)\rangle_{\rm pair} 
&= 
\frac{q\mu}{\displaystyle 2  \cosh(\mu L/2)}\sum_{j=0,1} (-1)^{j+\lfloor \eta_j(x)\rfloor}
\nonumber\\
&\qquad\times
\sinh\left[(\{\eta_j(x)\}-1/2) \mu L\right]\,.
\end{align} 
The electric field that corresponds to (\ref{eq:Theta-x-2-probes})
is plotted in FIG.~\ref{figure:charge-density-electric-field-graph}(b).
Performing the summation in~(\ref{eq:electric-field-interval}), we obtain the explicit expression
\begin{align}
 \left\langle F_{01}  \right\rangle_{\rm pair}& = 
 \frac{qg}{\displaystyle 2  \cosh(\mu L/2)}\sum_{j=0,1} (-1)^{j+\lfloor \eta_j(x)\rfloor}
\nonumber \\ 
&\quad\quad\quad\quad
\times 
 \cosh\left[(\{\eta_j(x)\}-1/2) \mu L\right]\,.
 \label{eq:electric-field-pair}
\end{align}

\paragraph{Behaviors near a boundary.}
We now consider the massless Schwinger model on a half-line~$[0,\infty)$ with $\Theta(x)=0$ and the boundary condition $\phi(x=0)=0$.
Let us begin with the energy density.
The regularized energy density on a half-line is obtained from (\ref{eq:energy-density-interval}) by sending $L$ to infinity:
$
\mathcal{E}^\text{half-line}_\Lambda(x):= \lim_{L\rightarrow \infty} \mathcal{E}_\Lambda(x) 
$
We define the renormalized energy density on a half-line as $\mathcal{E}_\text{half-line} := \lim_{\Lambda\rightarrow\infty} \left(
\mathcal{E}^\text{half-line}_\Lambda(x)
-
\mathcal{E}^\text{line}_\Lambda
\right) $.
We find
\begin{equation}\label{eq:energy-density-half-line}
\mathcal{E}_\text{half-line} =
 -\frac{\mu^2}{2\pi} K_{0}(2\mu x) \,.
\end{equation}
The modified Bessel function $K_0(z)$ has the asymptotics
\begin{equation}
K_0(z)= \left\{
\begin{matrix}
-\log(z/2) -\gamma +\mathcal{O}(z)& {\rm for}\ z\sim0 \,,\\
\displaystyle \sqrt{\frac{\pi}{2 z}}  {\rm e}^{-z} \left( 1+ \mathcal{O}(1/z)\right) & {\rm for}\ z\gg 1 \,.
\end{matrix}
\right.
\end{equation}
Thus the energy density diverges logarithmically near the boundary and decays exponentially away from it.
From
\begin{equation}
\lim_{L\rightarrow\infty} \lambda(x)=\exp\left[
\int_0^\infty dk \frac{\cos(2k x)}{\sqrt{\mu^2+k^2}} 
\right] 
=
{\rm e}^{K_0(2\mu x)} \,.
\end{equation}
we get for the chiral condensate 
\begin{equation}\label{eq:condensate-half-line}
\langle \bar\psi\psi  (x)\rangle_\text{half-line} 
=- \displaystyle\frac{{\rm e}^\gamma}{2\pi^{3/2}} g
{\rm e}^{K_0(2\mu x)} \,,
\end{equation}
which is the result obtained in~\cite{Kao:2001qn}.
The condensate diverges as $\langle \bar\psi\psi  (x)\rangle =\mathcal{O}(x^{-1/2})$ near the boundary and decays exponentially away from it.
The charge density and the electric field simply vanish an half-line for $\Theta(x)=0$ and the boundary condition $\phi(x=0)=0$.

\paragraph{Behaviors near a probe charge.}
Let us consider the system with a single probe charge $q$ at $x=x_0$ on an infinite line, which is represented by the position-dependent theta angle
\begin{equation}\label{eq:Theta-single-W}
\Theta_\text{probe}(x) :=\left \{
\begin{array}{clc}
0& \text{ for }x <x_0 \,,\\
\vspace{-4mm} \\
2\pi q & \text{ for } x>x_0 \,.
\end{array}
\right. 
\end{equation}
By taking an appropriate limit of~(\ref{eq:energy-density-interval}) or by repeating the steps leading to~(\ref{eq:energy-density-interval}), we obtain the energy density (renormalized by subtracting the value without a probe)
\begin{equation}\label{eq:energy-density-probe}
\mathcal{E}_\text{probe}(x) =
   \frac{\pi  }{4}   q^2\mu^2 {\rm e}^{-2\mu|x-x_0|} \,. 
\end{equation}
In a similar manner one can obtain expressions for other local observables.
We obtain, as in Appendix~D of~\cite{q-sim-confinement},
\begin{align}
& \langle \bar\psi\psi(x)\rangle_\text{probe} =  -  \frac{{\rm e}^\gamma g}{2\pi^{3/2}}
\nonumber\\
& \qquad
\times
\left\{
\begin{array}{cl}
\cos\left[\pi q {\rm e}^{-\mu(x_0-x)} \right] & {\rm for}\ x<x_0 \,, \\
\cos\left[2\pi q - \pi q {\rm e}^{-\mu(x-x_0)} \right]  &{\rm for}\ x>x_0  \,.
\end{array}
\right.
\label{eq:condensate-probe}
\end{align}
We also have
\begin{equation}\label{eq:charge-density-probe}
\langle \bar\psi\gamma^0\psi(x)\rangle_\text{probe}= -\frac{q}{2}\mu  {\rm e}^{-\mu|x-x_0|} \,,
\end{equation}
which previously appeared in~(4.12) of~\cite{Iso:1988zi}.
Integrating this, one obtains
\begin{equation}\label{eq:electric-field-probe}
\langle F_{01}\rangle_\text{probe} = \frac{q}{2}g \, {\rm sgn}(x-x_0) {\rm e}^{-\mu |x-x_0|} \,.
\end{equation}

\paragraph{Behaviors near a boundary charge.}

We now consider the massless Schwinger model on a half-line~$[0,\infty)$ with $\Theta(x)=0$ and the boundary condition 
\begin{equation}\label{eq:boundary-charge-def}
\phi(x=0) =\sqrt2 w_0 =:q/2
\end{equation}
 (or equivalently $\Theta(x)=\sqrt\pi q$ and $\phi(0)=0$).
This can be obtained by setting $w_0=q/2$ and taking the limit $L\rightarrow\infty$.
We find that the energy density is the sum of~(\ref{eq:energy-density-half-line}) and~(\ref{eq:energy-density-probe}) while the charge density and the electric field are respectively given by~(\ref{eq:charge-density-probe}) and~(\ref{eq:electric-field-probe}) (all with $x_0=0$).
The chiral condensate is given by
\begin{equation}\label{eq:condensate-boundary-charge}
\langle \bar\psi\psi(x)\rangle_\text{bdry char.} =  -  \frac{{\rm e}^\gamma g}{2\pi^{3/2}} {\rm e}^{K_0(2\mu x)} \cos\left(\pi q e^{-\mu x}\right) \,.
\end{equation}

\section{Lattice Schwinger model with open boundary conditions}\label{sec:lattice}

\subsection{Fermion versus spin models on a lattice}\label{sec:lattice-formulation}

Let us turn to the Kogut-Susskind lattice formulation of the Schwinger model~\cite{Kogut:1974ag,Banks:1975gq} with a position-dependent theta angle.
We wil
and follow the conventions of~\cite{q-sim-confinement}.

We consider a one-dimensional spatial lattice with $N$ sites, labeled by integers $n=0,1,\ldots,N-1$.\footnote{We will see that the behavior of the model depends strongly on whether $N$ is even or odd.
}
The two components $\psi_u(x)$ and $\psi_d(x)$ of the Dirac fermion  $\psi=(\psi_u,\psi_d)^T$ are replaced by the staggered fermion $\chi_n$ at the even and odd sites with $x=na$ respectively, according to the correspondence
\begin{equation}\label{eq:psi-chi}
\begin{array}{cc}
\psi_u(x) & \\
\psi_d(x) &
\end{array}
\quad
\longleftrightarrow
\quad
\frac{\chi_n}{\sqrt{2a}}
\begin{array}{ccc}
& n:& \text{even} \\
 & n: &\text{odd.} 
\end{array}
\end{equation}
On the $n$-th link, which connects the sites $n$ and $n+1$, we introduce the link variables $U_n$ and $L_n$  satisfying $U_n^\dagger = U_n^{-1}$, $L_n^\dagger = L_n$ according to the correspondence
\begin{equation} \label{correspondence:A1-Un_Pi-Ln}
{\rm e}^{-{\rm i} ag A^1 (x)} \longleftrightarrow U_n  \,,
 \quad
- \Pi(x)/g
 \longleftrightarrow L_n \,.
\end{equation}
These operators satisfy canonical (anti-)commutation relations,
among which the non-trivial ones are 
\begin{equation}
\{ \chi_m,\chi_n^\dagger \} = \delta_{mn}\,, 
\quad
[U_m ,L_n] =\delta_{mn}U_m \,.
\end{equation}
We also introduce the lattice version $\vartheta_n$ of the position-dependent theta angle on the $n$-th link:
\begin{equation}
\Theta(x) \longleftrightarrow \vartheta_n \,.
\end{equation}
The Hamiltonian of the lattice theory is
\begin{equation}
  \label{eq:fermion-Hamiltonian}
  \begin{aligned}
&
 \hspace{-2mm}
H_{\rm lattice}  =
 \frac{g^2a}{2} 
   \hspace{-1mm}
    \sum_{n=0}^{N-2}
      \hspace{-1mm}
      \left( L_n +\frac{\vartheta_n}{2\pi}  \right)^2 
 \hspace{-2mm}
 -\frac{\rm i}{2a} 
  \hspace{-1mm}
 \sum_{n=0}^{N-2} \Big( \chi_n^\dag U_n \chi_{n+1} 
 \\
 &\quad
 -\chi_{n+1}^\dagger U_n^\dag \chi_{n} \Big) 
  +m\sum_{n=0}^{N-1} (-1)^n \chi_n^\dagger \chi_n\,,
  \end{aligned}
\end{equation}
which is the direct counterpart of 
(\ref{eq:Hamiltonian-density}).

There is a relation between $N$ and the fermion boundary conditions.
As in~(\ref{eq:psi-chi}) we identify $\psi_u$ ($\psi_d$) with $\chi_{\rm even}$ ($\chi_{\rm odd}$).
Since  $n$ in $\chi_n$ runs from $0$ to $N-1$, we effectively have $\chi_{-1}=\chi_N=0$.
Thus we have $\psi_d=0$ on the left and $\psi_u=0$ ($\psi_d=0$) on the right for $N$ even (odd), namely there is a correspondence, leading to to the NS (R) boundary condition.%
\footnote{%
We also checked that the DMRG computation of the spectra of the XY model, which is equivalent to the free fermion model via the Jordan-Wigner transformation for open b.c.'s,  reproduces the expected spectra of the continuum Dirac fermion obeying the NS (R) boundary conditions for large $N$ even (odd).
}

As in the continuum theory, the physical Hilbert space is obtained by the Gauss law constraint.
The standard choice~\cite{Hamer:1997dx} of the Gauss law constraint is%
\footnote{Here, the presence of external charges is accounted for, as in~(\ref{eq:Theta-x}), by the position-dependent theta angle $\vartheta_n$ in~(\ref{eq:fermion-Hamiltonian}).
Cf.~(\ref{eq:Gauss_lattice-q}).
}
\begin{equation}\label{eq:Gauss_lattice}
G_n^\text{standard} := L_n -L_{n-1} -  \chi_n^\dag \chi_n +\frac{1-(-1)^n}{2}  =0 \,.
\end{equation}
We impose the boundary condition
$L_{-1}=0$ and fix the gauge $U_n =1$
to eliminate $(L_n ,U_n )$.
  The term $(-1)^n/2$ in
  ~(\ref{eq:Gauss_lattice}) represents a site-dependent background charge.  
In the bulk, the spatially averaged background charge density vanishes in the continuum limit, but we will see that there remains a non-trivial localized charge on a boundary and induces a background electric field.

We convert the fermions into spin variables by the Jordan-Wigner transformation~\cite{Jordan:1928wi}
\begin{equation}
 \chi_n = \frac{X_n-{\rm i} Y_n}{2}\prod_{i=0}^{n-1}(- {\rm i} Z_i) \,,
\end{equation}
where $X_n,Y_n,Z_n$ respectively denote the Pauli matrices $\sigma_x,\sigma_y,\sigma_z$ associated with the $n$-th site.
Besides the theta angle, $g$ and $m$ as in Section~\ref{sec:continuous},
the lattice introduces the lattice spacing~$a$ as an extra parameter.
The length~$L$ of the spatial interval is given by $L=(N-1)a$.
The Gauss law constraint reads
\begin{equation}\label{eq:Gauss-spin}
0 = L_n -L_{n-1} - \frac{Z_n+(-1)^n}{2} \,.
\end{equation}
We solve this with the boundary condition
\begin{equation}\label{eq:Lminus1}
L_{-1}=0 \,.
\end{equation}
The Hamiltonian in terms of the spin variables is
\begin{equation}\label{eq:spin-hamiltonian}
\begin{aligned}
& H_{\rm spin} = 
\frac{g^2 a}{2}\sum_{n=0}^{N-2} \left[\sum_{i=0}^{n}\frac{Z_i + (-1)^i}{2}+\frac{\vartheta_n}{2\pi}\right]^2 
 \\
 & +
\frac{1}{4a}
\hspace{-0.5mm}
\sum_{n=0}^{N-2}\left(X_nX_{n+1}+Y_{n}Y_{n+1}\right)
+\frac{m}{2}
\hspace{-0.5mm}
\sum_{n=0}^{N-1}(-1)^n Z_n 
 \,.
\end{aligned}
\end{equation}
Note the structural similarity between~(\ref{eq:Hamiltonian-density-bosonized}) and (\ref{eq:spin-hamiltonian}).

We have the following correspondence for the local observables of the continuum theory and the spin model.
\begin{align}
\mathcal{H}(x)\big|_{m=0}
& \ \longleftrightarrow\   
\begin{array}{ccc}
\displaystyle  
\frac{g^2 }{2} \left[ \sum_{i=0}^{n}\frac{Z_i + (-1)^i}{2}+\frac{\vartheta_n}{2\pi}\right]^2 
    \\
    \\
\displaystyle  
\quad +   \frac{1}{4a^2}(X_nX_{n+1}+Y_{n}Y_{n+1})
\,,
\end{array}
\\
\bar\psi\gamma^0\psi(x)
& \ \longleftrightarrow\ 
\frac{1}{4a} (Z_n+Z_{n+1})\,, 
\\
\bar\psi\psi(x)
&\ \longleftrightarrow\ 
\frac{(-1)^n}{4a } (Z_n - Z_{n+1})
   \,,
\\
 F_{01}
&\ \longleftrightarrow\ 
g 
 \sum_{i=0}^n \frac{Z_i+(-1)^i}{2} +  \frac{\vartheta_n}{2\pi} 
 \,.
\end{align}
The quantities on the left and right hand sides are for a continuous theory and a lattice model respectively, requiring  renormalization (normal ordering) in the former.


We will often consider the particular form of the position-dependent theta angle corresponding to probe charges $\pm q$ located at the sites $n=\hat{\ell}_0$  $n=(\hat{\ell}_0+\hat{\ell})$:
\begin{align}
\label{eq:theta-pair}
(\vartheta_\text{pair})_n 
:= \left\{
 \begin{array}{cc}
  2\pi q +\theta_0, & \hat{\ell}_0 \le n < \hat{\ell}_0+\hat{\ell},\\
  \theta_0, & \text{otherwise}.
 \end{array}
 \right.
\end{align}

\subsection{Spin lattice versus bosonized continuum models}
\label{sec:spin-vs-bosonized}

Let us compare the Gauss law constraints~(\ref{eq:Gauss-spin}) and (\ref{eq:Gauss-solution-bosonized})  in the spin and bosonized formulations, respectively.
The correspondence
\begin{equation} \label{eq:Ln-F01}
\frac{1}{g}F_{01} - \frac{1}{2\pi}\Theta(x) 
\quad
\longleftrightarrow  
\quad
L_n 
\end{equation}
in~(\ref{correspondence:A1-Un_Pi-Ln})
suggests the correspondence 
\begin{equation}\label{eq:phi-Z}
\phi(x)
\quad
\longleftrightarrow  
\quad
\frac{\sqrt\pi}{2}\sum_{i=0}^n \left(Z_i+(-1)^i\right) =:\phi_n \,.
\end{equation}
The operator $\phi_n$ rotates the $X_jY_j$ planes for $j\leq n$.
The comparison of~(\ref{eq:Hamiltonian-density-bosonized}) and~(\ref{eq:spin-hamiltonian})  suggests the correspondence
\begin{align}\label{eq:correspondence-of-kinetic-terms}
&
\frac12(\Pi_\phi)^2+\frac12(\partial_x\phi)^2 + {\rm const.}
\longleftrightarrow
\nonumber\\
&\qquad\qquad\qquad
\frac{1}{4a^2}\left(X_nX_{n+1}+Y_{n}Y_{n+1}\right)
 =: h_n \,.
\end{align}
Taking the commutators of the both sides of~(\ref{eq:phi-Z}) and~(\ref{eq:correspondence-of-kinetic-terms}) gives another correspondence
\begin{equation}\label{eq:correspondence-of-Pi}
\Pi_\phi
\longleftrightarrow 
 \frac{\sqrt\pi}{4a} (Y_n X_{n+1} - X_n Y_{n+1}) =: \pi_n \,,
\end{equation}
where the expression on the right arises from $[\phi_m,h_n] = ({\rm i}/a) \delta_{mn}\pi_n $.
We note the commutation relation
\begin{equation}\label{eq:phi-pi-comm}
[\phi_m, \pi_n ] = - \pi {\rm i} \delta_{mn} h_n  \,.
\end{equation}
This reduces to the canonical commutation relation between $\phi(x)$ and $\Pi_\phi(x)$  in the continuum limit because the density of the kinetic term diverges as $ -1/(\pi a^2)$%
\footnote{%
The divergence can be computed by the free fermion because it is not affected by $g$ or $m$, which only appears as $ga$ or $ma$.
}
 so that we can replace $h_n$ by $ -1/(\pi a)$ in~(\ref{eq:phi-pi-comm}).

The lattice Schwinger model described in Section~\ref{sec:lattice-formulation} should correspond, in the continuum limit, to
specific values of $w_0$ and $w_1$ in~(\ref{def:w0-w1}).
In the appendix of~\cite{q-sim-confinement}, it was argued that
\begin{equation}\label{eq:w01-values}
w_0 = \frac{1}{4} \,,  \quad w_1 =\frac{Q}{2}+\frac{1}{4}\,, 
\quad
Q:=\sum_{n=0}^{N-1}Z_n \,.
\end{equation}
The charge~$Q$ is conserved and can be treated as a $c$-number within a fixed charge sector.
%
In fact, if the value of $\nu$ in~(\ref{eq:nu-def})
 is $\nu_0$ ($\nu_1$) at $x=0$ ($x=L$), 
 we have%
 \footnote{%
This should follow from the bosonization rules $\psi_{\rm L(R)}\sim e^{{\rm i} \phi_{\rm L(R)}}$, where $\phi_{\rm L(R)}$ is the normalized left-(right-)moving part of $\phi$.
We checked it by comparing the explicit cylinder partition functions.}
\begin{equation}\label{eq:nu-w}
\nu_1-\nu_0 = w_1 - w_0 
\quad \text{ mod } \mathbb{Z} \,.   
\end{equation}
The relations~(\ref{eq:w01-values}) and~(\ref{eq:nu-w}) are non-trivially consistent with the correspondence between with parity of $N$ and the fermion boundary conditions found in Section~\ref{sec:lattice-formulation}.
We will also explicitly confirm the identification~(\ref{eq:w01-values}) by comparing the charge densities computed by DMRG and by bosonization.
We note that the eigenvalue of $Q$ is even (odd) if $N$ is even (odd).
Thus the winding number  $w_1-w_0$ is an integer (a half-integer) if $N$ is even (odd).
To summarize, we have the  correspondence%
\footnote{%
For a similar correspondence in the case of periodic XY models,
see Section 5.2.2 
of~\cite{fradkin_2013}.
}
\begin{equation}\label{eq:bc-corresp}
\hspace{-1mm}
\begin{array}{ccccc}
 N \text{ even} &\longleftrightarrow & \text{integer winding} &  \longleftrightarrow &\text{NS b.c.,}
 \\
 N \text{ odd} &\longleftrightarrow & \text{half-integer winding} &  \longleftrightarrow & \text{R b.c.}
 \end{array}
\end{equation}
\begin{figure*}[t]
\begin{center}
\begin{tabular}{cc}
\includegraphics[scale=0.9]{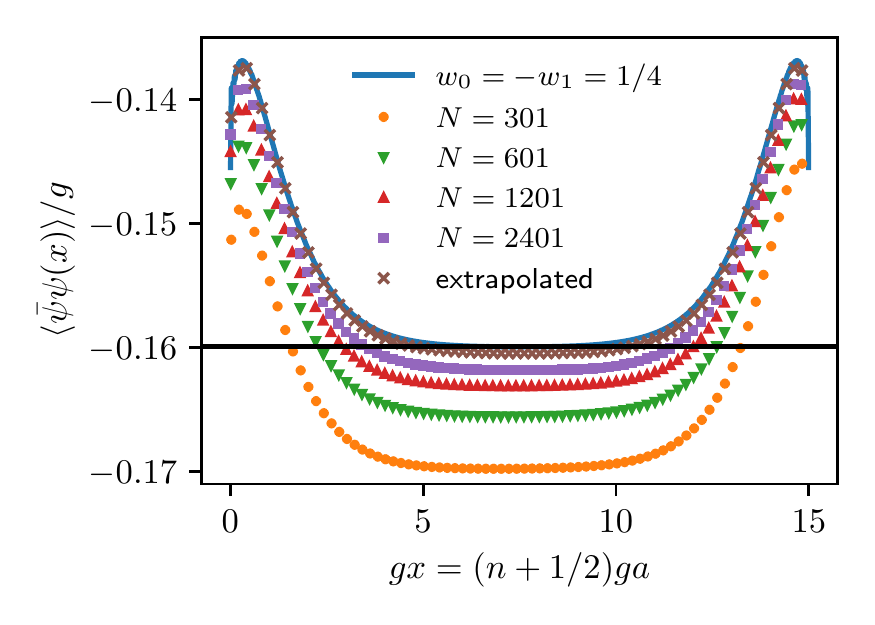}
&
\includegraphics[scale=0.9]{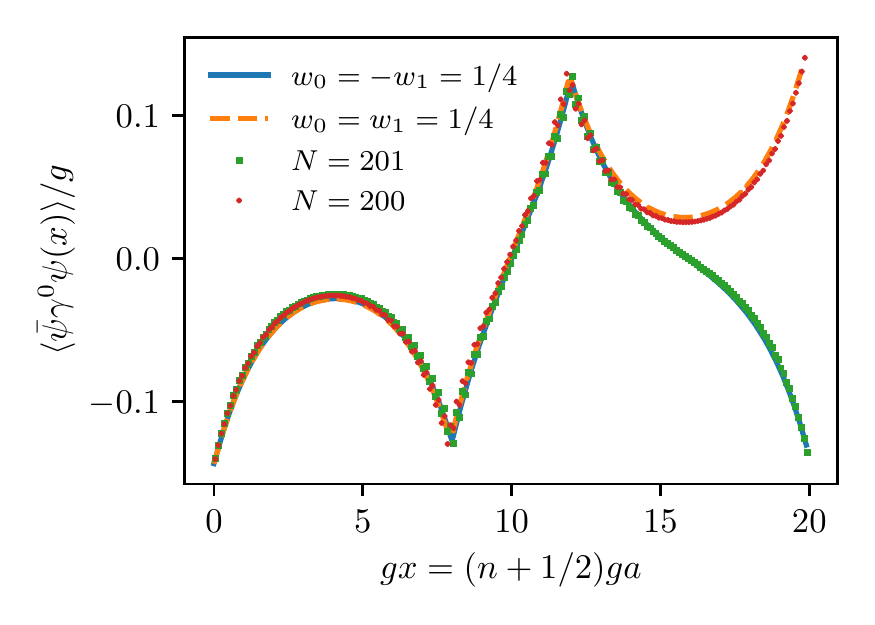}
\\
\hspace{16mm} (a)&\hspace{17mm}(b)
\end{tabular}
\end{center}
\caption{(a) Chiral condensate for $L= (N-1)a=15g^{-1}$ and $q=0$ (no probe charge), $m=0$, and $\theta_0=0$.
Plots of $(4ga)^{-1}(-1)^n\langle Z_n-Z_{n+1}\rangle$ computed by DMRG and $\langle\bar\psi\psi(x)\rangle/g $ computed analytically by the formula~(\ref{eq:chiral-condensate-Theta}) are shown.
We also plot the results of extrapolation to $N=\infty$ ($a=0$) obtained by fitting the data for $N\in\{301,601,1201,2401\}$ by a quadratic function of $1/N$.
For better visibility, only a subset of the values of $n$ is used.
(b) Charge density for a pair of charges with $g\ell=4$,  $q=0.5$, and $\theta_0=m=0$.
Plots of $(4ga)^{-1} \langle (Z_n+Z_{n+1})\rangle $ computed by DMRG and $\langle\bar\psi\gamma^0(x)\rangle/g$ computed analytically by (\ref{eq:charge-density-interval}) and (\ref{eq:Theta-x-2-probes}) are shown.
The precise length of the interval is $L=(N-1)a$ with $ga=0.1$.
}\label{figure:charge-density_condensate_DMRG}
\end{figure*}

\subsection{Comparison of DMRG and analytic results}
\label{sec:DMRG-analytic}


Here we compare the DMRG results based on the spin formulation in Section~\ref{sec:lattice} and the analytic results based on bosonization in Section~\ref{sec:bosonization}.
For 
our implementation of DMRG, we used the ITensor library~\cite{itensor}.
See~\cite{Honda:2022edn} for a related study.

For the chiral condensate $\propto \langle (-1)^n Z_n\rangle$, we plot the DMRG results including the extrapolated values and the analytical results in FIG.~\ref{figure:charge-density_condensate_DMRG}(b) for the case with no probe charges.
After extrapolation, the DMRG and analytical results match well.

For the charge density, we plot the DMRG and analytic results in FIG.~\ref{figure:charge-density_condensate_DMRG}(a).  We see that they agree very well.
This gives strong evidence for the identification~(\ref{eq:w01-values}).
Near the right boundary, the charge density profile is identical to that near a probe of charge $-1/2$ for $N$ even and $+1/2$ for $N$ odd.
We note that the parameters $w_0$ and $w_1$ parametrizing the Dirichlet boundary conditions for $\phi$ in~\ref{def:w0-w1} are related to the boundary charges $q_L$ and $q_R$ on the left and right boundaries as
\begin{equation}\label{eq:w-q}
w_0 = \frac{q_L}{2} \,,
\quad
w_1 = - \frac{q_R}{2} \,,
\end{equation}
generalizing (\ref{eq:boundary-charge-def}).
Therefore, the charges on the boundaries are half-integral, signifying charge fractionalization.
Indeed, for $N$ both even and odd, the charge density near left boundary has a spatial profile identical to the charge density near a probe charge $+1/2$.

\subsection{DMRG with a modified Gauss law constraint}
\label{sec:modified-gauss}

\begin{figure*}[t]
\begin{center}
\begin{tabular}{cc}
\includegraphics[scale=0.9]{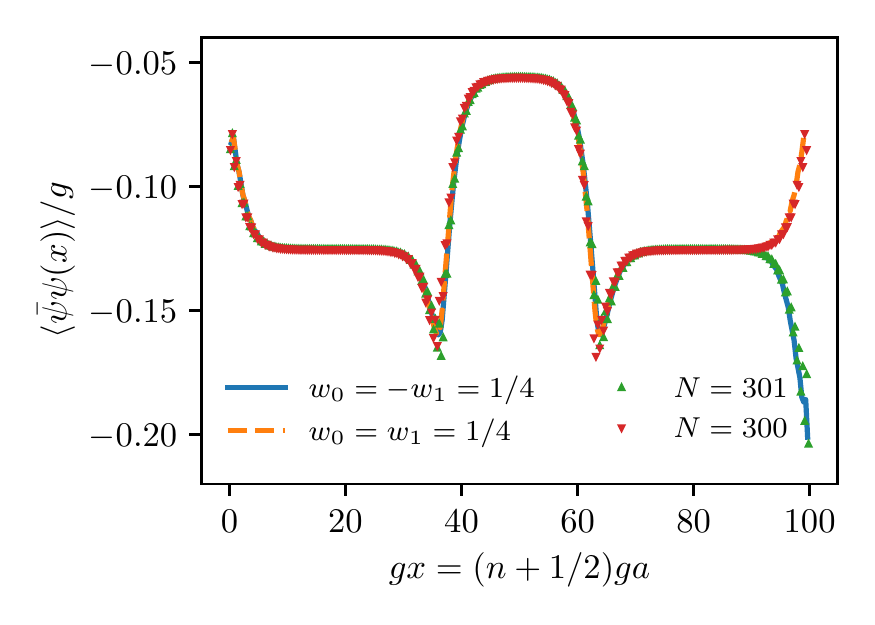}
&
\includegraphics[scale=0.9]{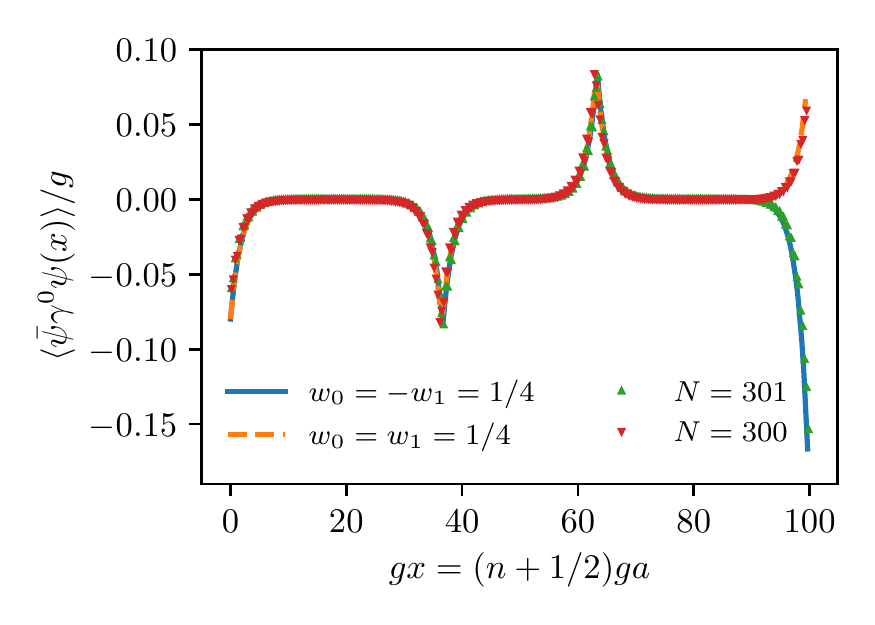}
\\
\hspace{16mm} (a)&\hspace{17mm}(b)
\end{tabular}
\end{center}
\caption{Plots of (a) the chiral condensate $\langle\bar\psi\psi(x)\rangle/g$ and (b) the charge density $\langle\bar\psi\gamma^0\psi(x)\rangle/g$.
The chiral condensate is computed analytically by (\ref{eq:chiral-condensate-Theta}), and by DMRG as $(4ga)^{-1}(-1)^n\langle Z_n-Z_{n+1}\rangle$.
The charge density is computed analytically by (\ref{eq:charge-density-interval}) and by DMRG as  $(4ga)^{-1}\langle Z_n+Z_{n+1}\rangle$.
The DMRG computation was done using (\ref{eq:theta-pair}) with $q=0.3$, $\theta_0=0.9$, $ga=1/3$, $\hat\ell=80$, $\hat\ell_0=\lfloor(N-\hat\ell-1)/2\rfloor$,  $m=0$, and  the values of $N$ indicated in the figure.
The analytic results shown as solid and dashed lines are computed for $\Theta(x) = \Theta_{(q,\theta_0-\pi/2)}(x)$ defined in (\ref{eq:Theta-x}) with $\ell =\hat\ell a$ and $\ell_0=\hat\ell_0 a$.
}\label{fig:Gauss-modified}
\end{figure*}

Above, we used
 the standard Gauss law constraint (\ref{eq:Gauss_lattice}), or equivalently (\ref{eq:Gauss-spin}), 
 for DMRG.
The constraint (\ref{eq:Gauss_lattice}) is chosen~\cite{Hamer:1997dx} so that it is satisfied by the ground state $|{\rm GS}_0\rangle$ in the ``strong coupling limit'' ($ga\rightarrow +\infty$ with $m/g^2a$ fixed)~\cite{Banks:1975gq} with vanishing $L_n$.
In terms of the fermion occupation numbers $\chi_n^\dagger \chi_n$, $|{\rm GS}_0\rangle$ corresponds to $|010101\ldots\rangle$.

In this subsection we consider a modified version of the Gauss law constraint~\cite{Berruto:1997jv}
\begin{equation}\label{eq:Gauss_lattice-modified}
\begin{aligned}
0 &= G_n^\text{modified} 
\\
&:= L_n -L_{n-1} -  \chi_n^\dag \chi_n +\frac{1}{2} =   L_n -L_{n-1} -  \frac{Z_n}{2} \,.
\end{aligned}
\end{equation}
Compared with the standard choice~(\ref{eq:Gauss_lattice}), we dropped the term $-(-1)^n/2$, which
 affects the boundary value of the scalar $\phi$, as argued in the appendix of~\cite{q-sim-confinement}.

If the periodic b.c. is chosen, as explained in Appendix B.4 of~\cite{Dempsey:2022nys}, this modification of the Gauss law constraint is equivalent, via a shift of $L_n$ by $(-1)^n/4$,%
\footnote{%
The corresponding manipulation for the open b.c. is~(\ref{eq:manip-bc}).
} 
to a shift of the mass parameter such that the theory with a vanishing shifted mass enjoys a discrete chiral symmetry and a faster convergence to the continuum limit.
While one has to allow $L_n$ to take non-integer values to satisfy the modified Gauss law~(\ref{eq:Gauss_lattice-modified}), one can require the shifted version to take integer values.
Solving the modified constraint with the boundary condition~$L_{-1}=0$ and fixing the gauge, we obtain the modified Hamiltonian
\begin{equation}\label{eq:spin-hamiltonian-modified}
\begin{aligned}
&H_{\rm modified} = 
\frac{g^2 a}{2}\sum_{n=0}^{N-2} \left[ \frac12\sum_{i=0}^{n}Z_i+\frac{\vartheta_n}{2\pi}\right]^2 
 \\
 &
\
+
\frac{1}{4a}
\hspace{-1mm}
\sum_{n=0}^{N-2}\left(X_nX_{n+1}+Y_{n}Y_{n+1}\right)
\hspace{-0.5mm}
+
\hspace{-0.5mm}
\frac{m}{2}
\hspace{-0.5mm}
\sum_{n=0}^{N-1}(-1)^n Z_n 
 \,.
\end{aligned}
\end{equation}

A direct calculation shows that
\begin{equation}\label{eq:manip-bc}
\begin{aligned}
&
 \left[ \frac12\sum_{i=0}^{n}Z_i+\frac{\vartheta_n}{2\pi}\right]^2 
\hspace{-2mm}
=
  \left[ \sum_{i=0}^{n} \frac{Z_i + (-1)^i}{2}
  +\frac{\vartheta_n - \frac{\pi}{2}}{2\pi}\right]^2 
    \\
  &\qquad \qquad
  - \sum_{i=0}^{N-1} \frac{(-1)^i}{8}  Z_i 
 - \frac{(-1)^N}{8} Q + \text{c-number} \,.
\end{aligned} 
\end{equation}
Comparing with~(\ref{eq:spin-hamiltonian}) we see that, within the fixed charge $Q$ sector, the modification~(\ref{eq:Gauss_lattice-modified}) of the Gauss law is equivalent to a shift of the mass $m \rightarrow m -  (g^2 a/8)$ \cite{Dempsey:2022nys} {\it and} a shift of the theta angle $\vartheta_n \rightarrow \vartheta_n - (\pi/2)$.
The latter shift would be further modified if we chose a boundary condition other than $L_{-1}=0$.

FIG.~\ref{fig:Gauss-modified} displays the profiles of the chiral condensate $\langle\bar\psi\psi(x)\rangle$ and the charge density $\langle\bar\psi\gamma^0\psi(x)\rangle$ computed by analytic formulas and DMRG for $m=0$.
Contrary to FIG.~\ref{figure:charge-density_condensate_DMRG}(b), extrapolation is unnecessary because the modification of the Gauss law, which is partially equivalent to the mass shift of~\cite{Dempsey:2022nys}, makes the convergence to the continuum limit much faster.

\section{Summary and discussion}\label{sec:conclusion}

In this work, we studied three formulations of the Schwinger model: the original fermionic formulation, the bosonized formulation, and the Hamiltonian lattice formulation.
We computed analytically physical observables in the ground state using the bosonized formulation and found excellent agreements with the DMRG computations in the lattice formulation.
We clarified the correspondence between boundary conditions in different formulations.
We studied a non-standard Gauss law constraint~(\ref{eq:Gauss_lattice-modified}) in the lattice formulation, and showed that it is equivalent to the mass shift of~\cite{Dempsey:2022nys} and a shift of the theta angle.
In accordance with \cite{Dempsey:2022nys}, we found that the modification of the Gauss law makes the convergence to the continuum limit faster.

As for future directions, it would  be interesting to re-derive our analytic results in the path integral formalism, along the line of~\cite{Sachs:1991en}.
It would also be worthwhile to establish the faster convergence more firmly by computing the precise difference between the lattice and continuum Hamiltonians.
This should be possible by classifying the potential counterterms to the local observables along the line of~\cite{Luscher:1996sc,Luscher:1998pe} that deals with the Euclidean path integral.
Finally, one should be able to perform DMRG in a similar manner to compute local observables in non-abelian lattice gauge theories in $1+1$ dimensions.

\section*{Acknowledgements}

The author thanks M.~Honda, E.~Itou, Y.~Kikuchi, and L.~Nagano for collaboration on~\cite{q-sim-confinement}, which motivated this work.
He also thanks Y.~Kikuchi for providing useful ITensor codes, based on which he wrote his own.
He is grateful to Y.~Tanizaki for pointing out an error in~\cite{q-sim-confinement}, whose correction was important in this work.
Some of the results in this paper were presented in the workshop ``Quantum computing for quantum field theories'' held at the Yukawa Institute for Theoretical Physics in January, 2021.
This work was supported in part by MEXT-JSPS Grant-in-Aid for Transformative Research Areas (A) ``Extreme Universe”, No. 21H05190.

\appendix

\section{Computation of the energy by the method of images}\label{app:images}

In this appendix we compute the ground state energy of the massless Schwinger model with probe charges, using the effective potential obtained in~\cite{Gross:1995bp}.

By integrating out the matter field and restricting to a static gauge field, the effective Lagrangian, on an infinite spatial line, is found to be
\begin{equation}\label{eq:Veff}
\mathcal{L}_{\rm eff}=\frac{1}{2}(\partial_1 A_0)^2 + \frac{\mu^2}{2}A_0^2 - \rho A_0\,,
\end{equation}
where we introduced the density~$\rho(x)$ of external charges.
For two charges $q_1$ and $q_2$ separated by distance $\ell$, $\rho(x)=q_1 g [\delta(x) + q_2 \delta(x-\ell)]$, the solution to the Euler-Lagrange equation
\begin{equation}\label{eq:A0-rho}
(-\partial_1^2   + \mu^2 )A_0 =\rho
\end{equation}
gives the two-body potential~\cite{Iso:1988zi}
\begin{equation}\label{eq:2-body-potential}
V_{q_1,q_2} (\ell)=  - \frac{\pi}{2}   q_1 q_2 \mu  \left(1-{\rm e}^{- \mu  \ell}\right)  \,.
\end{equation}

To compute the energy on an interval~$[0,L]$,
we extend the domain of the charge density $\rho(x)$ from 
to $(-\infty,\infty)$ as an even periodic function of period $2L$, $\rho(-x)=\rho(x)$, $\rho(x+2L)=\rho(x)$.
We solve~(\ref{eq:A0-rho}) for $A_0$ using the Green's function~$G(x)={\rm e}^{-\mu|x|}/2\mu$ and substitute~$A_0$ to~(\ref{eq:A0-rho}) with the integration range $[0,L]$.
The energy given by
\begin{equation}\label{eq:energy-rho-A0}
E = - \int_0^L dx \mathcal{L}_{\rm eff} = \frac{1}{2}\int_0^L dx \rho(x) A_0(x) 
\end{equation}
can be evaluated by summing the two-point potential between 1) the probe charges in the interval $[0,L]$, and 2) the probe charges in the interval and image charges.

\begin{figure}[t]
\begin{center}
\includegraphics[scale=.3]{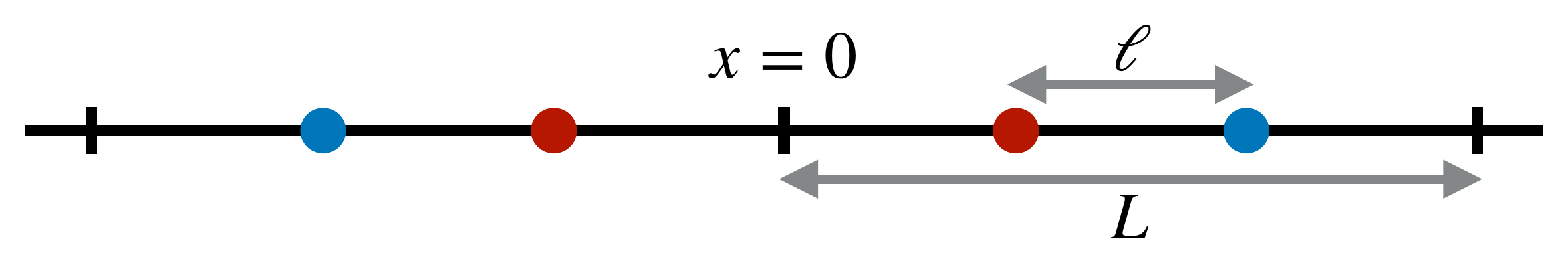}
\end{center}
\caption{%
Charge distribution~$\rho_{\rm pair}(x)$ extended as an even periodic function.
}\label{figure:probes-mirror-1}
\end{figure}

As an example let us consider the charge distribution~$\rho_{\rm pair}(x)=q\delta(x-\ell_0)-q\delta(x-\ell_0-\ell)$ with $\ell_0=(L-\ell)/2$ for  $0<x<L$, shown in FIG.~\ref{fig:probes-interval}.
We extend $\rho_{\rm pair}(x)$ to an even periodic function, 
which is depicted in FIG.~\ref{figure:probes-mirror-1}.
The energy~(\ref{eq:energy-rho-A0}) reproduces~(\ref{eq:E-pair}).

\begin{figure}[t]
\begin{center}
\includegraphics[scale=.3]{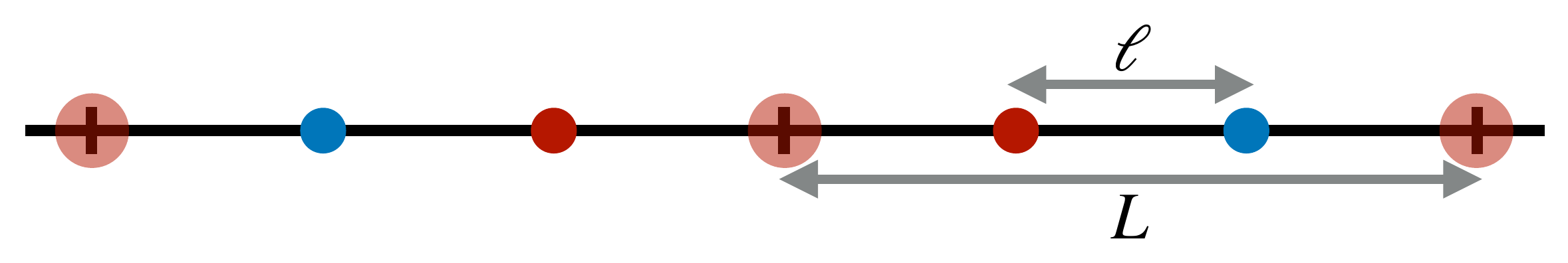}
\end{center}
\caption{%
Charge distribution with boundary charges of the same sign and magnitude added.
}\label{figure:probes-mirror-3}
\end{figure}

As another example, let us modify the set-up in the previous paragraph by adding charges $q_L$ and $q_R$ to the left and right boundaries, respectively.
We are interested in the $q$-dependent part of the energy.
To compute it, we sum the two-point potentials between 1) the probe charges in the interval $[0,L]$,  2) the probe charges in the interval and their image charges, and 3) the probe charges in the interval and the boundary charges including their images.
The two-point potential between boundary charges is $q$-independent and we drop it.
Compared with the previous paragraph, the new contribution is from 3):
\begin{equation}\label{eq:Delta-E-pair}
\Delta E_\text{pair} = \frac{\pi  \mu  }{2} q(q_L-q_R)
\frac{{\rm e}^{ - \frac{\mu}{2}   (L-\ell)}-{\rm e}^{-\frac{\mu}{2}   (L+\ell)} 
   }{ 1+{\rm e}^{-\mu  L}} \,.
\end{equation}
This vanishes if $q_L=q_R$.
See FIG.~\ref{figure:probes-mirror-3}.

For $\Theta =\Theta_{\rm pair}$ in~(\ref{eq:Theta-x-2-probes}) and general $w_0$ and $w_1$ in~(\ref{def:w0-w1}), the ground state energy computed from~\eqref{eq:Hamiltonian-bosonized}) turns out to be
\begin{equation}
{(\rm \ref{eq:E-pair})}
+ \pi \mu q (w_0+w_1)  \frac{\sinh(\mu\ell/2)}{\cosh(\mu L/2)} \,.
\end{equation} 
This is consistent with~(\ref{eq:Delta-E-pair}) by the relations in~(\ref{eq:w-q}), {\it i.e.}, $w_0=q_L/2$ and $w_1=- q_R/2$.
For $w_0=-w_1=1/4$, $\Delta E_\text{pair} =0$.
This result appeared and was used in~\cite{q-sim-confinement}.
We checked that the DMRG computation of the ground state energy with large $N$  agrees well (as a function of $\ell$) with~$ E_\text{pair} + \Delta E_\text{pair}$ for $(q_L,q_R)=(1/2,-1/2)$ if $N$ is even, and for $(q_L,q_R)=(1/2,1/2)$ if $N$ is odd.

\section{Exact one-form symmetries in lattice QCDs in $1+1$ dimensions}
\label{sec:one-form}

In this appendix, we show that the general {\it lattice} QCD in the Kogut-Susskind formulation~\cite{Kogut:1974ag} enjoys an {\it exact} one-form~$C_0$ symmetry, where $C_0$ consists of the elements of the center of the gauge group~$G$ under which the matter fermions are invariant.
The presence of such a center 1-form symmetry well-known in the continuum limit~\cite{Gaiotto:2014kfa} and is also known for the charge-$q$ lattice Schwinger model~\cite{Dempsey:2022nys}.
As the  discussion for the general lattice QCD is rather abstract, we begin with the charge-$q$ Schwinger model, which can be understood more intuitively.

The charge-${\rm q}$ Schwinger model, {\it i.e.}, 
the $U(1)$ gauge theory with a single Dirac fermion of charge ${\rm q}$, 
has attracted attention in recent years.
See, e.g., \cite{Honda:2021ovk,Cherman:2021nox}.
As the defining action we take 
\begin{equation}\label{eq:Schwinger-action-q}
\begin{aligned}
S
&=\int d^2x \Big[ -\frac{1}{4} F_{\mu\nu} F^{\mu\nu} +\frac{g\vartheta}{4\pi} \epsilon_{\mu\nu} F^{\mu\nu}
\\
&
\quad +{\rm i}\bar{\psi}\gamma^\mu (\partial_\mu +{\rm i} {\rm q} g A_\mu ) \psi -m\bar{\psi}\psi  \Big]
\\
&\quad\quad
 +  \sum_p q_p  g\int dt A_t(x_p)+\text{boundary terms}
 \,.
\end{aligned}
\end{equation}
Unlike in Section~\ref{sec:continuous}, we define probe charges using couplings separate from the theta angle.
We also take $q_p$ to be integers so that the corresponding Wilson lines are genuine line operators rather than boundaries of topological surface operators.
The bosonized Lagrangian is
\begin{equation}
\begin{aligned}
\mathcal{L}&=
-\frac14 F_{\mu\nu}F^{\mu\nu} 
+\frac{g}{4\pi}
\vartheta
\epsilon^{\mu\nu}F_{\mu\nu} 
 + \frac{{\rm q}g}{\sqrt\pi} \epsilon^{\mu\nu} A_\mu \partial_\nu \phi 
\\
&\quad
+ \frac12 \partial_\mu\phi\partial^\mu\phi 
+ m g \frac{{\rm e}^\gamma}{2\pi^{3/2}}  \cos(2\sqrt\pi\phi) 
\\
&\quad\quad
+\sum_p q_p g  \delta(x-x_p) A_t(x)
\,.
\end{aligned}
\end{equation}
The Gauss law constraint reads
\begin{equation}\label{eq:Gauss-bosonized-q}
\partial_1\left(F_{01}  - \frac{g}{2\pi} (\vartheta +2{\rm q}\sqrt\pi \phi) \right) =\sum_p q_p   g\delta(x-x_p)   \,.
\end{equation}
%
The theory possesses a $\mathbb{Z}_{\rm q}$ one-form symmetry, whose generator can be expressed in the bosonized form~\cite{Honda:2021ovk,Cherman:2021nox}
\begin{equation} \label{eq:Vq-bosonized}
\begin{aligned}
V_{\rm q} &:= \exp\left[\frac{2\pi {\rm i}}{{\rm q}g} \left(F_{01} -\frac{g}{2\pi}\vartheta - \frac{{\rm q}g}{\sqrt\pi}\phi\right)\right]
\nonumber\\
&
=
\exp\left( \sum_p \frac{2\pi{\rm i}}{\rm q} q_p H_{\rm step}(x-x_p)\right)  
\,.
\end{aligned}
\end{equation}
This is piecewise constant  as a topological operator should be, and labels the distinct decomposed sectors of the theory called ``universes"~\cite{Pantev:2005rh,Pantev:2005wj,Pantev:2005zs,Hellerman:2010fv,Tanizaki:2019rbk}.

The corresponding Hamiltonian of the lattice theory in the presence of probe charges is
\begin{equation} \label{eq:fermion-Hamiltonian-q-probe}
\begin{aligned}
H_{\rm lattice}  &= J \sum_{n=0}^{N-2} \left( L_n +\frac{\vartheta}{2\pi}  \right)^2 
 - {\rm i}w \sum_{n=0}^{N-2} \Big[ \chi_n^\dag (U_n)^{\rm q} \chi_{n+1} 
 \\
 &\qquad  
  -\chi_{n+1}^\dagger (U_n^\dagger)^{\rm q} \chi_{n} \Big]
+m\sum_{n=0}^{N-1} (-1)^n \chi_n^\dagger \chi_n \,.
\end{aligned}
\nonumber
\end{equation}
Again, we work in a formulation slightly different from Section~\ref{sec:lattice} and~\cite{Honda:2021ovk} and implement the effects of probe charges by adding the corresponding terms in the Gauss law constraint
\begin{equation}\label{eq:Gauss_lattice-q}
 L_n -L_{n-1}  +{\rm q}\left[
 \hspace{-0.5mm}
 -   \chi_n^\dag \chi_n +\frac{1-(-1)^n}{2} \right]
  \hspace{-0.5mm}
=
 \hspace{-0.5mm}
 \sum_p q_p \delta_{n n_p} \,.
\end{equation}
In terms of spin variables we have
\begin{equation}\label{eq:Gauss_lattice-q-spin}
L_n -L_{n-1} - {\rm q} \frac{Z_n+(-1)^n}{2}  = \sum_p q_p \delta_{n n_p} \,.
\end{equation}
The lattice generator of the $\mathbb{Z}_{\rm q}$ one-form symmetry, corresponding to~(\ref{eq:Vq-bosonized}), is~\cite{Dempsey:2022nys} 
\begin{equation}\label{eq:one-form-gen-q}
\begin{aligned}
V_{\rm q} 
&= \exp\left[
\frac{2\pi {\rm i}}{\rm q}
\left(
L_n- \frac{\rm q}{2} \sum_{i=0}^n \left( Z_i + (-1)^i \right)
\right)
\right] 
\\
&=
\exp\left(
\frac{2\pi {\rm i}}{\rm q}
L_n
\right)
\,,
\end{aligned}
\end{equation}
where we used the correspondences~(\ref{eq:Ln-F01}) and~(\ref{eq:phi-Z}) and the fact that $Z_i + (-1)^i$ vanishes mod 2.
As in the continuum case, the Gauss law~(\ref{eq:Gauss_lattice-q-spin}) implies that $V_{\rm q} $ acting on a physical state is almost constant as a function of the position but gets multiplied by a phase as one crosses probe charges $q_p$ (temporal Wilson lines) at $n=n_p$.
This means, by the Wick rotation and the exchange of space and time, that $V_{\rm q} $ obeys the expected commutation relations with the Wilson lines.%
\footnote{%
In~\cite{Honda:2022edn} it was demonstrated that the complexified chiral condensate $\langle \bar\psi e^{{\rm i}\gamma_5}\psi\rangle$ flows in the IR to the topological operator $V_q$.
}

We now turn 
 to an arbitrary  Hamiltonian lattice gauge theory with a general gauge group $G$ and a fermion in representation $\rho$~\cite{Kogut:1974ag}.
For $G$ we only require that it is compact: it can be non-abelian, discrete, a product, a quotient, or something more complicated.
The Lie algebra $\mathfrak{g}$ of $G$ decomposes into simple Lie algebras
\begin{equation}
\mathfrak{g} = \bigoplus_b \mathfrak{g}_b 
\,.
\end{equation}
The maximal torus $T$ of $G$ has the Lie algebra $\mathfrak{t} = \bigoplus_b \mathfrak{t}_b$, where $\mathfrak{g}$ and $\mathfrak{g}_b$ are the Cartan subalgebras of  $\mathfrak{g}$ and  $\mathfrak{g}_b$, respectively.
The center~$C$ of $G$ has the Lie algebra $\mathfrak{c}=\bigoplus_i \mathfrak{c}_i$, where $\mathfrak{c}_i\simeq \mathbb{R}$ is the Lie algebra of the ``$U(1)$ factor" labeled by $i$.
In general $\rho$ is reducible: 
\begin{equation}
\rho = \bigoplus_f\rho_f \,,
\end{equation}
where $\rho_f$ is an irreducible representation of $G$.
Again
we consider a one-dimensional lattice with sites labeled by $n=0,1,\ldots N-1$.
The Hilbert space of the theory is the tensor product of the local Hilbert spaces associated with sites $n\in\{0,\ldots,N-1\}$ and links $n\in\{0,\ldots,N-2\}$.
On each site~$n$, we have a fermion Fock space, possibly tensored with the representation space for a probe charge.
The fermionic Fock space is generated by the fermion $\chi_n=(\chi^f_n)_f $ in representation $\rho=\bigoplus_f \rho_f$ and its hermitian conjugate.
In addition, if we place a probe charge in representation ${\rm R}_p$ on site $n_p$, we tensor the Fock space with the representation space~${\rm V}_p$ of ${\rm R}_p$.
On each link~$n$ we have the space of square-integrable functions on $G$.
The total Hilbert space is thus of the form
\begin{equation}
\mathcal{H}_\text{total} = \mathcal{H}_\text{fermion} \otimes \mathcal{H}_\text{gauge} \otimes \mathcal{H}_\text{probe} 
\,.
\end{equation}
The Hamiltonian takes the form%
\footnote{%
We do not include a kinetic term for the discrete part of the gauge group.
Thus if the whole gauge group is discrete and there is no matter, the gauge theory is topological.
}
\begin{widetext}
\begin{equation}
H_{\rm lattice}  
= \sum_b J_b \sum_{n=0}^{N-2}{\rm tr} \left( L^{b}_n +\frac{\delta_{b i}\vartheta_i}{2\pi}  \right)^2 
 - {\rm i}w \sum_{n=0}^{N-2} \big( \chi_n^\dag  \rho(g_n) \chi_{n+1} 
 -\chi_{n+1}^\dagger \rho(g_n)^\dagger\chi_{n} \big)  
 +
 \sum_f
 m_f \sum_{n=0}^{N-1} (-1)^n (\chi^f_n)^\dagger \chi^f_n \,,
\end{equation}
\end{widetext}
where $w=1/2a$,  $J_\alpha = g_b^2a/2 $, $g_b$ is the coupling constant for $\mathfrak{g}_b$, $\vartheta_i$ is the theta angle for $\mathfrak{c}_i$, and $m_f$ is the mass for the fermion labeled by~$f$.
The trace~${\rm tr}$ is taken in a faithful irreducible representation of $G$, in which $g_n$ is represented by a unitary matrix $U_n$.
(The $\mathfrak{g}_b$ part of) the ``left' canonical momentum $L_n  = \oplus_b L^b_n$ conjugate to $g_n\in G$ can be expanded as $L_n = L_{\alpha n} T^\alpha $ and obeys (in our sign convention) the canonical commutation relation
\begin{equation}
[g_m, L_{\alpha n}] = \delta_{mn}T_\alpha  g_m \,,
\end{equation}
where $T_\alpha = \kappa_{\alpha\beta} T^\beta$, the matrix $\kappa_{\alpha\beta}$ is the inverse of $\kappa^{\alpha\beta} = {\rm tr}(T^\alpha T^\beta)$, which is a Killing form of $\mathfrak{g}$.
Let us define $R_n:=g_n^{-1} L_n g_n$.
Then one can show that $R_{\alpha n}$ defined by $R_n = R_{\alpha n} T^\alpha $ satisfies the commutation relation
\begin{equation}
[g_m, R_{\alpha n}] = \delta_{mn}  g_m T_\alpha \,.
\end{equation}
The group~$\mathcal{G}$ of gauge transformations is the product
of copies of $G$, each associated with a site $n$.
The gauge transformation $h_n\in G$ on site $n$ acts as
\begin{equation}\label{eq:gauge-hn}
\begin{aligned}
& \chi_n \rightarrow \rho(h_n) \chi_n \,,
 \quad
 g_n \rightarrow h_n g_n h_{n+1}^{-1} \,,
\\
& L_n \rightarrow h_n L_n h_{n}^{-1} \,,
 \quad
 R_n \rightarrow h_{n+1} R_n h_{n+1}^{-1} \,,
\end{aligned}
\end{equation}
and leaves the Hamiltonian invariant and the canonical commutation relations invariant.
For the continuous part of the gauge group, the Gauss law constraints are
\begin{equation} \label{eq:Gauss-law-general}
L_n^\alpha - R^\alpha_{n-1}  - \chi_n^\dagger T^\alpha \chi_n + \frac{1-(-1)^n}{2} {\rm tr}(T^\alpha) = \sum_p \delta_{nn_p} T^\alpha_p  \,,
\nonumber
\end{equation}
where $T^\alpha_p $ are the generators in the representations $R_p$ for probe charges.
It is possible to consider, as in~(\ref{eq:Gauss_lattice-modified}), the modified version of the Gauss law constraint where the term containing $(-1)^n$ is dropped.

If the gauge group $G$ contains as a factor the cyclic group $\mathbb{Z}_d = \{ 
{\rm e}^{(2\pi{\rm i}/d) j}
\, |\, j=0,1,\ldots,d-1\} $,
 on each link there exist operators
 $Z_n$ and $X_n$ such that $Z_m X_n= \exp[(2\pi{\rm i}/d)\delta_{mn}]X_nZ_m $, $Z_n^d=X_n^d=1$.
The Gauss law constraint takes the group form
\begin{equation}
\begin{aligned}
&
\hspace{-1mm}
X_n X_{n-1}^{-1} \exp\left[\frac{2\pi {\rm i}}{d}\left(
\hspace{-0.5mm}
- \chi_n^\dagger D \chi_n +\frac{1-(-1)^n}{2}{\rm tr}\,D
\hspace{-0.5mm}
\right)\right]
\\
&\qquad\qquad\qquad
= \exp\left[\frac{2\pi{\rm i}}{d} \sum_s q_s \delta_{nn_s}\right]\,,
\end{aligned}
\end{equation}
where~$D$ is a diagonal matrix of $\mathbb{Z}_d$ charges (integers modulo $d$) for the fermion, and $q_s$ are the  $\mathbb{Z}_d$ charges of the probes at $n=n_s$.

For a general gauge group (including non-abelian discrete groups such as the dihedral group~$D_4$) 
instead of imposing the Gauss law constraint in terms of operators,
we can simply project the total Hilbert space $\mathcal{H}_\text{total}$ onto the physical Hilbert space $\mathcal{H}_\text{phys}$, which is the subspace of $\mathcal{H}_\text{total}$ invariant under the group~$\mathcal{G}$ of gauge transformations~\cite{Lamm:2019bik}.

To study one-form symmetries, let $C_0$ consist of the elements of the center~$C$ (of the gauge group~$G$) under which the fermion $\chi_n$ is invariant.
Since $C_0$ is abelian and compact, it is of the form $C_0 = U(1)^M \times \Gamma$, where $\Gamma$ is a product of cyclic groups.
On each site~$n$ and for $c\in C_0$, let us consider the operator~${\rm Gauge}_n(c)$ 
 implementing the gauge transformation corresponding to
$c^{-1}$.%
\footnote{%
Here we use $c^{-1}$ instead of $c$ to be consistent with earlier definitions of one-form symmetry generators.
}
It is of the form
\begin{equation}
{\rm Gauge}_n(c)= {\rm Left}_n(c^{-1}) {\rm Right}_{n-1}(c){\rm R}_p(c^{-1})^{\delta_{n n_p}} \,,
\nonumber
\end{equation}
 where ${\rm Left}_n(h)$ (resp. ${\rm Right}_n(h)$) is the operator corresponding to the left (resp. right) action of $h$ on the copy of $G$ on link $n$.
The appearance of $ {\rm Left}_n(c^{-1})$ and $ {\rm Right}_{n-1}(c)$ can be understood from~(\ref{eq:gauge-hn}).
The operator ${\rm R}_p(c^{-1})^{\delta_{n n_p}}$ represents the action of $c^{-1}$ on the representation space ${\rm V}_p$ for the probe~$p$.
Because~$c$ is in the center, in fact we have $ {\rm Left}_n(c^{-1}) =  {\rm Right}_n(c^{-1}) =: V_n(c)$.
On the physical Hilbert space~$\mathcal{H}_\text{phys}$, which is invariant under gauge transformations, we have the equality $V_n(c) V_{n-1}(c)^{-1} {\rm R}_p(c^{-1})^{\delta_{n n_p}}=1$ or equivalently
\begin{equation} \label{eq:one-form-gen-general}
V_n(c)  = V_{n-1}(c) {\rm R}_p(c)^{\delta_{n n_p}} \,.
\end{equation}
Since $c$ is in the center and  ${\rm R}_p$ is an irreducible representation, ${\rm R}_p(c) = \exp\left[{\rm i} \alpha_{R_p}(c)\right]$ is in fact a c-number corresponding to the charge under $C_0$.
Equation~(\ref{eq:one-form-gen-general}) establishes that $V_n(c)$ is the generator of the one-form symmetry for $C_0$.
It is constant between probe charges, and obeys the expected commutation relation between Wilson line operators $W_{\rm R} = {\rm Tr}_{\rm R} P \exp\left({\rm i} \oint A\right)$:
\begin{equation}
W_{\rm R} V(c) 
=
  {\rm e}^{{\rm i} \alpha_R(c)}
  V(c) W_{\rm R}  \,,
\end{equation}
which we rewrote as an operator relation via a Wick rotation and a rotation in the Euclidean spacetime.

\bibliography{refs}

\end{document}